\documentstyle[preprint,aps]{revtex}

\begin{document}
\draft
\title{Generalized Schwinger proper time method for Dirac operator
 with dynamical chiral symmetry breaking}

\author{Qin Lu$^a$,   Hua Yang$^{a,c}$, Qing Wang$^{a,b}$,}

\address{$^a$Department of Physics, Tsinghua University, Beijing 100084, China
\footnote{Mailing address}\\
$^b$Institute of Theoretical Physics, Academia Sinica, Beijing 100080, China\\
$^c$Science School, Information Engineering University, Zhengzhou 450004}
\date{Nov 4, 2002}

\maketitle
\begin{abstract}
Schwinger proper time method is generalized for the calculation of real part of determinant and coincidence limit of inverse for Dirac operator with
 dynamical chiral symmetry breaking caused by momentum
dependent fermion self energy $\Sigma(k^2)$. The obtained series
generalizes the heat kernel  expansion for hard fermion mass.
\end{abstract}

\bigskip
PACS number(s): 03.65.Db, 11.10Ef, 11.30.Rd, 12.39.Fe

\section{Introduction}

In a renormalizable quantum field theory at 4-dimensional Euclidean
space time, the local interaction of a fermion system with external fields
is bilinear in fermion fields and realized in terms of Dirac operator $D$,
\begin{eqnarray}
&&D\equiv\nabla\!\!\!\! /\;-s+ip\gamma_5\hspace{2cm}
\nabla_{\mu}\;\equiv \partial_{\mu}-iv_{\mu}-ia_{\mu}\gamma_5
=-\nabla_{\mu}^{\dagger}\;\;,\label{nabladef}
\end{eqnarray}
where $s$, $p$, $v_{\mu}$, $a_{\mu}$ are hermitian
fields fermion coupled with and $\gamma$ matrix is
hermitian. We
 have decomposed these fields in terms of their Lorentz structures
\footnote{In general, there should be tensor field, but for simplicity of
discussion, in this paper, we donot consider the tensor field.}.
In terms of Dirac spinor $\psi$ and $\overline{\psi}$,
the action of renormalizable interaction is
\begin{eqnarray}
\int d^4x~~\overline{\psi}D\psi\label{psiint0}
\end{eqnarray}
This action is invariant under following local chiral transformations
\begin{eqnarray}
\psi(x)\rightarrow\psi'(x)&=&[V_L(x)P_L+V_R(x)P_R]\psi(x)\nonumber\\
J(x)\rightarrow J'(x)&=&
[V_L(x)P_R+V_R(x)P_L][J(x)+i\partial\!\!\! /\;_x][V_L^{\dagger}(x)P_L
+V_R^{\dagger}(x)P_R]\label{chiral}
\end{eqnarray}
with $V_L(x)$ and $V_R(x)$ be left and right chiral rotation matrices and
project operator $P_{^R_L}\equiv\frac{1}{2}(1\pm\gamma_5)$.  The
external field $J(x)$ is defined as
\begin{eqnarray}
J(x)=-iv\!\!\! /\;(x)-ia\!\!\! /\;(x)\gamma_5-s(x)+ip(x)\gamma_5
\end{eqnarray}
Suppose we have $N_f$ Dirac spinors in our fermion system, this local chiral
symmetry then is $U_L(N_f)\otimes U_R(N_f)$.

The contribution of action (\ref{psiint0}) to a path integral of 4-
dimension space time quantum field theory can be written as
\begin{eqnarray}
\int{\cal D}\overline{\psi}{\cal D}\psi~e^{
\int d^4x~~[\overline{\psi}D\psi+\overline{I}\psi+\overline{\psi}I]}
=e^{{\rm Tr}\ln D-\int d^4x\int d^4y\overline{I}(x)
D^{-1}(x,y)I(y)}\label{path}
\end{eqnarray}
where $\overline{I}$ and $I$ are external sources of the fermion system. $D$ is
 Dirac operaor which in general depend on external fields. Due to
bilinear property of fermion interaction, we have integrated out fermions in
the path integral  (\ref{path}) and found that
the contribution of fermion to the path
integral is realized through fermion determinant Trln$D$ and propagator
$D^{-1}(x,y)=\langle x|D^{-1}|y\rangle$ in presence of external
 fields. Once these two objects are obtained,
 the contribution of fermion system to
 a 4-dim space time renormalizable field theory is known.

In practice, for the physical needs and regularization of infrared divergence,
 a constant bare mass $m$ is usually added into the theory which is equivalent
  to extract out from the scalar field $s$
  a condensation part $s\rightarrow s-m$.
With this quark mass term, interaction action (\ref{psiint0}) become
\begin{eqnarray}
\int d^4x~~\overline{\psi}(D+m)\psi\label{psiint}
\end{eqnarray}
and two ingredients of fermion system in (\ref{path}) become
logarithm of fermion determinant Trln$(D+m)$ and fermion
propagator $(D+m)^{-1}(x,y)$.

Adding in quark mass term explicitly violate original chiral symmetry from
$U_L(N_f)\otimes U_R(N_f)$ to $U_V(N_f)$, whose transformation is in fact
requiring the original left handed and right handed rotations be same
 $V_L(x)=V_R(x)\equiv h^{\dagger}(x)$.

With constant
 quark mass term, a way to keep the theory still be chiral symmetric is
nonlinear realization of the theory. i.e.,
 we need to introduce into theory a local field $\Omega(x)$
 which  transform nonlinearly  under local chiral symmetry (\ref{chiral}) as
\begin{eqnarray}
\Omega(x)\rightarrow\Omega'(x)=h^{\dagger}(x)\Omega(x)V_L^{\dagger}(x)=
V_R(x)\Omega(x)h(x)\label{Omega}
\end{eqnarray}
and rotated external fields
\begin{eqnarray}
J_{\Omega}(x)&=&[\Omega(x)P_R+\Omega^{\dagger}(x)P_L]
~[J(x)+i\partial\!\!\!\! /\;]
~[\Omega(x)P_R+\Omega^{\dagger}(x)P_L]\nonumber\\
&\equiv&-s_{\Omega}(x) +ip_{\Omega}(x)\gamma_5 +v\!\!\! /\;_{\Omega}(x)
+a\!\!\! /\;_{\Omega}(x)\gamma_5
\label{JOmegadef}
\end{eqnarray}
in which $h(x)$ is a transformation matrix for induced hidden local symmetry
$U_V(N_f)$.
Replacing the external fields in $D+m$ in
(\ref{psiint}) with $D_{\Omega}+m$, i.e.
\begin{eqnarray}
D+m\rightarrow D_{\Omega}+m
\end{eqnarray}
with
$D_{\Omega}\equiv\nabla\!\!\!\! /\;_{\Omega}-s_{\Omega}
+ip_{\Omega}\gamma_5$,
$\nabla^{\mu}_{\Omega}\;\equiv \partial^{\mu}-iv^{\mu}_{\Omega}
-ia^{\mu}_{\Omega}\gamma_5$.
One can easily check that under local  chiral symmetry
$U_L(N_f)\otimes U_R(N_f)$ transformation (\ref{chiral}) and (\ref{Omega}),
\begin{eqnarray}
J_{\Omega}(x)&\rightarrow&J_{\Omega}'(x)=h^{\dagger}(x)
[J_{\Omega}(x)+i\partial\!\!\!\! /\;_x]h(x)\label{hidden}\\
D_{\Omega}+m
&\rightarrow&
D_{\Omega}'+m
=h^{\dagger}(x)
[D_{\Omega}+m]h(x)\label{hidden1}
\end{eqnarray}
i.e., they are covariant quantities. In terms of rotated quantities,
 local  chiral symmetry $U_L(N_f)\otimes U_R(N_f)$ is nonlinearly realized
 through hidden symmetry $U_V(N_f)$ by $h(x)$.
 Once our formulation keep this local hidden  symmetry $U_V(N_f)$, it
 automatically keep
 original  local  chiral symmetry $U_L(N_f)\otimes U_R(N_f)$. If our
  formulation  keep the local symmetry $U_V(N_f)$, it then keep
 local  chiral symmetry $U_L(N_f)\otimes U_R(N_f)$ in terms of $\Omega$ field.
So with constant
 quark mass term, to keep  local  chiral symmetry $U_L(N_f)\otimes
U_R(N_f)$ in the theory, we only need  to replace original sources with the
rotated sources. In this sense, one can easily generalize the theory with
 symmetry $U_V(N_f)$ to the case of symmetry $U_L(N_f)\otimes U_R(N_f)$
 in terms of $\Omega$ field. We in this paper will limit ourselves with symmetry
 $U_V(N_f)$ and unrotated sources.

 Beyond constant quark mass, in the literature, there is discussion on how
to generalize the constant fermion mass to a matrix \cite{matrix}
to respect further breaking of the $U_V(N_f)$ symmetry in the fermion system.
In this work, we are
interested in the generalization in another direction--
 the case of dynamical violating the chiral symmetry. Within context of
 dynamical chiral symmetry breaking (DCSB),
 the most general situation is not the appearance of a hard fermion mass $m$,
 but a momentum $k^2$
 dependent fermion self energy  $\Sigma(k^2)$. In fact, in the quantum field
 theory, even
 the chiral symmetry is broken by a momentum independent hard mass term
  at leading order of calculation, include in high order quantum corrections,
 the hard mass will be replaced by momentum dependent fermion self energy.
 Taking hard fermion mass often cause extra
 ultraviolet divergence, instead, momentum dependent fermion self
  energy usually will suppress the ultraviolet divergence. For example,
  in the case of massless QCD, the fermion
 self energy damping at least as $1/k^2$ at ultraviolet momentum region,
 if we take a hard fermion mass to substitute self energy,
 we will over estimate its contribution to physics and cause extra divergence
 which usually lead many physical fine tuning problems.
  To respect the momentum dependence of fermion self energy in the theory,
   in this paper, we generalize the conventional Schwinger
    proper time formulation \cite{Schwinger} to
   deal with corresponding fermion determinant and propagator. For simplicity
   and as the first step of research in this direction
 we only discuss the real part of fermion determinant
 known conventionally as effective action and part of coincidence limit of
 fermion propagator related to fermion pair condensation.
The imaginary part of fermion determinant  related to anomaly and remaining
nonlocal part of fermion propagator will be discussed else where.

This paper is organized as follows: In section II, we review the conventional
 Schwinger proper time formulation for the real part of fermion determinant
  and coincidence limit of fermion propagator. In section III, we discuss how to generalize the formulation from a hard mass
   case to momentum dependent dynamical fermion self energy.
   Section IV is responsible for the calculation
of the real part of fermion determinant and
  section V is responsible for the calculation
of the coincidence limit of fermion propagator in which the tensor structure is
ignored in our calculation. Section VI is summary and discussion.

\section{Review of Schwinger proper time formulation for Dirac operator with a
hard fermion mass}

Schwinger proper time formulation \cite{Schwinger,Seely} is a kind
of regularization method which can covariantly regularize the
ultraviolet divergence of Feynman loop diagrams. It is especially
effective in dealing with the  problems related to chiral gauge
theories, in which appearance of
 $\gamma_5$ matrix prohibits the naive
 use of conventional dimensional regularization scheme.
The most important application of Schwinger proper time formulation is
to calculate fermion determinant which plays a key role in Feynman loop
diagrams from the discussion of last section.

The real part of fermion determinant is related to logarithm of a hermitian
operator $(D^{\dagger}+m)(D+m)$
\begin{eqnarray}
{\rm Re}\ln{\rm Det}(D+m)&=&\frac{1}{2}
{\rm Tr}\ln[(D^{\dagger}+m)(D+m)]
\end{eqnarray}
Take
\begin{eqnarray}
E-\nabla^2=D^{\dagger}D+D^{\dagger}m+mD
\end{eqnarray}
and with help of following relation for a
matrix $(D^{\dagger}+m)(D+m)=E-\nabla^2+m^2$,
\begin{eqnarray}
\int_{\frac{1}{\Lambda^2}}^{\infty}\frac{e^{-(D^{\dagger}+m)(D+m)
\tau}}{\tau}d\tau&=&
-{\rm Ei}(-\frac{(D^{\dagger}+m)(D+m)}{\Lambda^2})\nonumber\\
&=&-\gamma-\ln(\frac{(D^{\dagger}+m)(D+m)}{\Lambda^2})
-{\displaystyle\sum_{n=1}^{\infty}}
\frac{(-1)^n}{n!n}(\frac{(D^{\dagger}+m)(D+m)}{\Lambda^2})^n
\end{eqnarray}
The real part of fermion determinant in conventional Schwinger proper time
formulation is
\begin{eqnarray}
{\rm Re}\ln{\rm Det}(D+m)&=&
\frac{1}{2}\lim_{\Lambda\rightarrow\infty}
\bigg[-\gamma+\ln\Lambda^2-
\int d^4x\int_{\frac{1}{\Lambda^2}}^{\infty}
\frac{d\tau}{\tau}~{\rm tr}~e^{-m^2\tau}
\langle x|e^{-\tau(E-\nabla^2)}|x\rangle\bigg]\label{lndetcal}
\end{eqnarray}
where tr is trace for internal symmetry indices, such as spinor,
flavor indices,etc. And
\begin{eqnarray}
&&(\nabla\!\!\!\! /\;)^{\dagger}\equiv -\partial\!\!\!/\;+iv\!\!\!
/\; -ia\!\!\! /\;\gamma_5\hspace{3cm} D^{\dagger}=\nabla\!\!\!\!
/\;^{\dagger}-s-ip\gamma_5\nonumber\\ &&E= -2ms-2ima\!\!\!
/\;\gamma_5 +\frac{i}{4}[\gamma^{\mu},\gamma^{\nu}]{\cal
R}_{\mu\nu} +\gamma_{\mu}d^{\mu}(s-ip\gamma_5)
+i\gamma^{\mu}[a_{\mu}\gamma_5(s-ip\gamma_5)+(s-ip\gamma_5)a_{\mu}\gamma_5]
\nonumber\\ &&\hspace{1cm}+s^2+p^2-[s,p]i\gamma_5\label{Edef}\\
 &&{\cal R}_{\mu\nu}\equiv i[\nabla_{\mu},\nabla_{\nu}]
=[d_{\mu}a_{\nu}-d_{\nu}a_{\mu}]\gamma_5+V_{\mu\nu}
-i[a_{\mu},a_{\nu}]\nonumber\\
&&d_{\mu}f=\partial_{\mu}f-i[v_{\mu},f]\hspace{2cm}
\hspace{2cm}V_{\mu\nu}
=\partial_{\mu}v_{\nu}-\partial_{\nu}v_{\mu}-i[v_{\mu},v_{\nu}]
\;\;.\nonumber
\end{eqnarray}
In Schwinger proper time formulation, the regularization is done at lower limit
 of proper time integration $\tau$. Since proper time is a parameter irrelevant
  to symmetry, this regularization has the feature of keeping symmetry of
    the system.

With help of standard Seely-DeWitt expansion \cite{Seely},
\begin{eqnarray}
\langle x|e^{-\tau(E-\nabla^2)}|x\rangle\label{Eexp}
&=&\frac{1}{16\pi^2}\bigg[\frac{1}{\tau^2}-\frac{E}{\tau}+\frac{1}{2}E^2
-\frac{1}{6}[\nabla_{\mu},[\nabla^{\mu},E]]
-\frac{1}{12}{\cal R}_{\mu\nu} {\cal R}^{\mu\nu}
-\frac{\tau}{6}E^3\label{Seely}\\
&&+\frac{\tau}{12}\{E[\nabla^{\mu},[\nabla_{\mu},E]]
+[\nabla^{\mu},[\nabla_{\mu},E]]E
+[\nabla^{\mu},E][\nabla_{\mu},E]\}+\frac{\tau^2}{24}E^4+\cdots\bigg]\;,
\nonumber
\end{eqnarray}
the proper time integration of $\tau$ in (\ref{lndetcal}) can be
finished, substitute (\ref{Edef}) into result formulae, we
 get the expansion of Re~lndet$(D+m)$ with hard fermion mass $m$,
\begin{eqnarray}
&&{\rm Re}\ln{\rm Det}(D+m)\nonumber\\ &&=
{\rm Re}\ln{\rm Det}(\partial\!\!\!\! /\;+m)
-\frac{1}{32\pi^2}\lim_{\Lambda\rightarrow\infty} \int d^4x
~{\rm tr}_f\bigg[8m[\Lambda^2+m^2(\ln\frac{m^2}{\Lambda^2}+\gamma-1)]s
\nonumber\\
&&\hspace{0.5cm}
-8m^2(\ln\frac{m^2}{\Lambda^2}+\gamma)a^2
-\frac{4}{3}[d_{\mu}a^{\mu}]^2
-\frac{2}{3}(\ln\frac{m^2}{\Lambda^2}+\gamma+1)
(d_{\mu}a_{\nu}-d_{\nu}a_{\mu})(d^{\mu}a^{\nu}-d^{\nu}a^{\mu})
\nonumber\\ &&\hspace{0.5cm}
-[\frac{4}{3}(\ln\frac{m^2}{\Lambda^2}+\gamma)+\frac{16}{3}]a^4
 +[\frac{4}{3}(\ln\frac{m^2}{\Lambda^2}+\gamma)
+\frac{8}{3}]a_{\mu}a_{\nu}a^{\mu}a^{\nu}
-4[\Lambda^2+m^2(3\ln\frac{m^2}{\Lambda^2}+3\gamma-1)]s^2
\nonumber\\ &&\hspace{0.5cm}
-4[\Lambda^2+m^2(\ln\frac{m^2}{\Lambda^2}+\gamma-1)]p^2
+(16\ln\frac{m^2}{\Lambda^2}+16\gamma+16)msa^2
-\frac{2}{3}(\ln\frac{m^2}{\Lambda^2}+\gamma)V_{\mu\nu}V^{\mu\nu}
\nonumber\\ &&\hspace{0.5cm}
+i[\frac{8}{3}(\ln\frac{m^2}{\Lambda^2}+\gamma)+\frac{16}{3}]a^{\mu}a^{\nu}V_{\mu\nu}
+8m(\ln\frac{m^2}{\Lambda^2}+\gamma)pd^{\mu}a_{\mu}
+\epsilon^{\sigma\rho\mu\nu}[-2
(\ln\frac{m^2}{\Lambda^2}+\gamma)V_{\sigma\rho}d_{\mu}a_{\nu}\nonumber\\
&&\hspace{0.5cm}
+4i(\ln\frac{m^2}{\Lambda^2}+\gamma+2)a_{\sigma}a_{\rho}d_{\mu}a_{\nu}]
+O(p^6)\bigg]\;\;,
\label{exp}
\end{eqnarray}
where tr$_f$ is the trace for flavor index and
we have isolated out the external fields independent term
\begin{eqnarray}
{\rm Re}\ln{\rm Det}(\partial\!\!\!\! /\;+m)
&=&-\frac{1}{2}\lim_{\Lambda\rightarrow\infty}
\int d^4x\int_{\frac{1}{\Lambda^2}}^{\infty}
\frac{d\tau}{\tau}~{\rm tr}~e^{-m^2\tau}
\langle x|e^{\tau\partial^2}|x\rangle\nonumber\\
&=&-\frac{1}{32\pi^2}\lim_{\Lambda\rightarrow\infty}
\int_{\frac{1}{\Lambda^2}}^{\infty}
\frac{d\tau}{\tau^3}
~e^{-m^2\tau}\int d^4x~{\rm tr}1
\nonumber\\
&=&-\frac{1}{32\pi^2}\lim_{\Lambda\rightarrow\infty}
[2\Lambda^4-4\Lambda^2m^2-m^4(2\ln\frac{m^2}{\Lambda^2}+2\gamma-1)]
\int d^4x{\rm tr}1
\end{eqnarray}
and the expansion terms in (\ref{exp})
are arranged according to its momentum power in which
 $\partial_{\mu}$, $v_{\mu}$, $a_{\mu}$ are treated as order $p$
and $s$, $p$  as order $p^2$ \cite{Gasser}. Hermitian matrix $\gamma_5$ is defined as $\gamma_5=-\gamma^0\gamma^1\gamma^2\gamma^3$ which leads relations
 tr$_l(\gamma_5\gamma^{\mu}\gamma^{\nu}\gamma^{\sigma}\gamma^{\rho})
 =-4\epsilon^{\mu\nu\sigma\rho}$ and $\gamma^{\mu}\gamma^{\sigma}\gamma^{\rho}=
 \gamma^{\mu}g^{\sigma\rho}-\gamma^{\sigma}g^{\mu\rho}+\gamma^{\rho}g^{\mu\sigma}-\epsilon_{\nu}^{~\mu\sigma\rho}\gamma^{\nu}\gamma_5$.

There are two ways to calculate coincidence limit of fermion propagator,
 one is direct perform functional differential with respect to corresponding
 external fields and the other is taking similar computation procedure as
 that for fermion determinant. We will see two methods give
  exactly same results.

First we consider the functional differential of fermion determinant. Use
identity
\begin{eqnarray}
&&\frac{\delta {\rm Tr}\ln(D+m)}{\delta J^{\sigma\rho}(x)}
=\int d^4yd^4z~(D+m)^{-1,\rho'\sigma'}(y,z)\frac{\delta D^{\sigma'\rho'}(z,y)}
{\delta J^{\sigma\rho}(x)}=(D+m)^{-1,\rho\sigma}(x,x)\nonumber\\
&&\frac{\delta {\rm Tr}\ln(D^{\dagger}+m)}{\delta J^{\dagger\sigma\rho}(x)}
=\int d^4yd^4z~(D^{\dagger}+m)^{-1,\rho'\sigma'}(y,z)
\frac{\delta D^{\dagger\sigma'\rho'}(z,y)}
{\delta J^{\dagger\sigma\rho}(x)}=(D^{\dagger}+m)^{-1,\rho\sigma}(x,x)
\label{derhard}
\end{eqnarray}
 where $\sigma=(a\xi),\rho=(b\zeta)$ ($a,b$ are flavor indices, $\xi,\zeta$
 are Lorentz indices).
The scalar part of  coincidence limit of fermion propagator then is
\begin{eqnarray}
&&-\frac{\delta {\rm ReTr}\ln(D+m)}{\delta s^{ab}(x)}
=-\frac{1}{2}\bigg[
\frac{\delta {\rm Tr}\ln(D+m)}{\delta s^{ab}(x)}
+\frac{\delta {\rm Tr}\ln(D^{\dagger}+m)}{\delta s^{ab}(x)}\bigg]\nonumber\\
&&=-\frac{1}{2}\int d^4y~\bigg[
\frac{\delta {\rm Tr}\ln(D+m)}
{\delta J^{\sigma\rho}(y)}
\frac{\delta J^{\sigma\rho}(y)}
{\delta s^{ab}(x)}
+\frac{\delta {\rm Tr}\ln(D^{\dagger}+m)}
{\delta J^{\dagger\sigma\rho}(y)}
\frac{\delta J^{\dagger\sigma\rho}(y)}
{\delta s^{ab}(x)}\bigg]\nonumber\\
&&=\frac{1}{2}(1)^{\xi\zeta}[(D+m)^{-1,(b\zeta)(a\xi)}(x,x)
+(D^{\dagger}+m)^{-1,(b\zeta)(a\xi)}(x,x)]
\nonumber\\
&&=\frac{1}{4\pi^2}\lim_{\Lambda\rightarrow\infty}
\bigg[m[\Lambda^2+m^2(\ln\frac{m^2}{\Lambda^2}+\gamma-1)]\delta^{ba}
-[\Lambda^2+m^2(3\ln\frac{m^2}{\Lambda^2}+3\gamma-1)]s^{ba}\nonumber\\
&&\hspace{0.5cm}+(2\ln\frac{m^2}{\Lambda^2}+2\gamma+2)m(a^2)^{ba}\bigg]
+O(p^4)\label{scalar}\;\;,
\end{eqnarray}
similarly the pseudo scalar part  is
\begin{eqnarray}
&&-i\frac{\delta {\rm ReTr}\ln(D+m)}{\delta p^{ab}(x)}
=-\frac{i}{2}\bigg[
\frac{\delta {\rm Tr}\ln(D+m)}{\delta p^{ab}(x)}
+\frac{\delta {\rm Tr}\ln(D^{\dagger}+m)}{\delta p^{ab}(x)}\bigg]\nonumber\\
&&=\frac{1}{4\pi^2}\lim_{\Lambda\rightarrow\infty}\bigg[
-i[\Lambda^2+m^2(\ln\frac{m^2}{\Lambda^2}+\gamma-1)]p^{ba}
+im(\ln\frac{m^2}{\Lambda^2}+\gamma)(d^{\mu}a_{\mu})^{ba}\bigg]
+O(p^4)\label{pseudo}\;\;,
\end{eqnarray}
the vector part  is
\begin{eqnarray}
&&i\frac{\delta {\rm ReTr}\ln(D+m)}{\delta v_{\mu}^{ab}(x)}
=\frac{i}{2}\bigg[
\frac{\delta {\rm Tr}\ln(D+m)}{\delta v_{\mu}^{ab}(x)}
+\frac{\delta {\rm Tr}\ln(D^{\dagger}+m)}{\delta v_{\mu}^{ab}(x)}\bigg]
\nonumber\\
&&=-\frac{i}{32\pi^2}\lim_{\Lambda\rightarrow\infty}
\bigg[-\frac{8i}{3}[d_{\nu}a^{\nu},a^{\mu}]
-\frac{8i}{3}(\ln\frac{m^2}{\Lambda^2}+\gamma+1)
[(d^{\mu}a^{\nu}-d^{\nu}a^{\mu}),a_{\nu}]
-\frac{8}{3}(\ln\frac{m^2}{\Lambda^2}+\gamma)d^{\nu}V^{\mu\nu}\nonumber\\
&&\hspace{0.5cm}
+i[\frac{8}{3}(\ln\frac{m^2}{\Lambda^2}+\gamma)+\frac{16}{3}]
d_{\nu}([a^{\mu},a^{\nu}])
+8im(\ln\frac{m^2}{\Lambda^2}+\gamma)[p,a^{\mu}]
+2i\epsilon^{\nu\sigma\rho\mu}(\ln\frac{m^2}{\Lambda^2}+\gamma)\bigg(2i
d^{\mu}d_{\sigma}a_{\rho}\nonumber\\
&&\hspace{0.5cm}+[a^{\mu},V_{\sigma\rho}]\bigg)\bigg]^{ba}+O(p^5)
\label{vector}\;\;,
\end{eqnarray}
the axial vector part  is
\begin{eqnarray}
&&-i\frac{\delta {\rm ReTr}\ln(D+m)}{\delta a_{\mu}^{ab}(x)}
=-\frac{i}{2}\bigg[
\frac{\delta {\rm Tr}\ln(D+m)}{\delta a_{\mu}^{ab}(x)}
+\frac{\delta {\rm Tr}\ln(D^{\dagger}+m)}{\delta a_{\mu}^{ab}(x)}\bigg]
\nonumber\\
&&=\frac{i}{4\pi^2}\lim_{\Lambda\rightarrow\infty}
\bigg[-2m^2(\ln\frac{m^2}{\Lambda^2}+\gamma)a^{\mu}
+\frac{1}{3}[d^{\mu}d^{\nu}a_{\nu}]
-\frac{1}{3}(\ln\frac{m^2}{\Lambda^2}+\gamma+1)
d^{\nu}(d^{\mu}a_{\nu}-d_{\nu}a^{\mu})\nonumber\\
&&\hspace{0.5cm}
-[\frac{1}{3}(\ln\frac{m^2}{\Lambda^2}+\gamma)+\frac{4}{3}]
(a^2a^{\mu}+a^{\mu}a^2)
 +[\frac{2}{3}(\ln\frac{m^2}{\Lambda^2}+\gamma)
+\frac{4}{3}]a_{\nu}a^{\mu}a^{\nu}\nonumber\\
&&\hspace{0.5cm}
+2(\ln\frac{m^2}{\Lambda^2}+\gamma+1)m(sa^{\mu}+a^{\mu}s)
+i[\frac{1}{3}(\ln\frac{m^2}{\Lambda^2}+\gamma)+\frac{2}{3}]
(a_{\nu}V^{\mu\nu}-V^{\mu\nu}a_{\nu})\nonumber\\
&&\hspace{0.5cm}
-m(\ln\frac{m^2}{\Lambda^2}+\gamma)d^{\mu}p
-\frac{1}{4}(\ln\frac{m^2}{\Lambda^2}+\gamma)d^{\mu}V_{\sigma\rho}
{\epsilon^{\nu\sigma\rho}}_{\mu}
\bigg]^{ba}
+O(p^5)\label{axial}
\;\;.
\end{eqnarray}
Since tensor fields are not introduced into the theory,
we cannot obtain the tensor part by performing functional differential.

We then take  the
second method, write propagator in presence of external fields as
\begin{eqnarray}
&&(D+m)^{-1}(x,x)\nonumber\\
&&=\bigg[[(D^{\dagger}+m)(D+m)]^{-1}(D^{\dagger}+m)\bigg](x,x)\nonumber\\
&&=\lim_{\Lambda\rightarrow\infty}\int_{\frac{1}{\Lambda^2}}^{\infty}
d\tau~e^{-m^2\tau}
\langle x|e^{-\tau(E-\nabla^2)}(D^{\dagger}+m)
|x\rangle\nonumber\\
&&=\lim_{\Lambda\rightarrow\infty}\int_{\frac{1}{\Lambda^2}}^{\infty}
d\tau~e^{-m^2\tau}
\bigg[-\langle x|e^{-\tau(E-\nabla^2)}\nabla\!\!\!\! /\;|x\rangle
+\langle x|e^{-\tau(E-\nabla^2)}|x\rangle[m-s(x)-ip(x)\gamma_5\nonumber\\
&&\hspace{0.5cm}-2ia\!\!\! /\;(x)\gamma_5]\bigg]
\label{propacal}
\end{eqnarray}
Similar as expansion (\ref{Eexp}), we have
\begin{eqnarray}
&&\langle x|e^{-\tau(E-\nabla^2)}\nabla_{\mu}|x\rangle
=\frac{1}{16\pi^2}\bigg[\frac{i}{6\tau}[\nabla^{\nu},{\cal R}_{\nu\mu}]
+\frac{1}{2\tau}[\nabla_{\mu},E]
-\frac{1}{3}E[\nabla_{\mu},E]-\frac{1}{6}[\nabla_{\mu},E]E\bigg]
+\cdots\label{E1exp}
\end{eqnarray}
Substitute (\ref{E1exp}) back to (\ref{propacal}) and finish the
integration of $\tau$,  we obtain the expansion of
$(D+m)^{-1}(x,x)$ with hard fermion mass $m$,
\begin{eqnarray}
&&(D+m)^{-1}(x,x)\nonumber\\
&&=\frac{1}{16\pi^2}\lim_{\Lambda\rightarrow\infty}
\bigg\{[\Lambda^2+m^2(\ln\frac{m^2}{\Lambda^2}+\gamma-1)](m-ip\gamma_5)
-2im^2(\ln\frac{m^2}{\Lambda^2}+\gamma)a\!\!\! /\;\gamma_5\nonumber\\
&&\hspace{0.5cm}-[\Lambda^2+m^2(3\ln\frac{m^2}{\Lambda^2}+3\gamma-1)]s
+im(\ln\frac{m^2}{\Lambda^2}+\gamma)d^{\mu}a_{\mu}\gamma_5
+2m(\ln\frac{m^2}{\Lambda^2}+\gamma+1)a^2
\nonumber\\
&&\hspace{0.5cm}+\bigg[
\frac{i}{3}(\ln\frac{m^2}{\Lambda^2}+\gamma)d^{\nu}V_{\mu\nu}
+m(\ln\frac{m^2}{\Lambda^2}+\gamma)(pa^{\mu}-a^{\mu}p)
-\frac{1}{3}[(d_{\nu}a^{\nu})a^{\mu}-a^{\mu}(d_{\nu}a^{\nu})]\nonumber\\
&&\hspace{0.5cm}+
\frac{1}{3}(\ln\frac{m^2}{\Lambda^2}+\gamma+1)[
a_{\nu}(d^{\mu}a^{\nu}-d^{\nu}a^{\mu})-(d^{\mu}a^{\nu}-d^{\nu}a^{\mu})a_{\nu}]
\nonumber\\
&&\hspace{0.5cm}
+\frac{1}{3}(\ln\frac{m^2}{\Lambda^2}+\gamma+2)
d^{\nu}(a_{\mu}a_{\nu}-a_{\nu}a_{\mu})
+\epsilon_{\mu}^{~\sigma\rho\nu}
[\frac{i}{2}(\ln\frac{m^2}{\Lambda^2}+\gamma)d^{\nu}d_{\sigma}a_{\rho}
\nonumber\\
&&\hspace{0.5cm}
+\frac{1}{4}(\ln\frac{m^2}{\Lambda^2}+\gamma-2)a^{\nu}V_{\sigma\rho}
-\frac{1}{4}(\ln\frac{m^2}{\Lambda^2}+\gamma+2)V_{\sigma\rho}a^{\nu}
+\frac{2i}{3}a_{\sigma}a_{\rho}a^{\nu}]\bigg]\gamma^{\mu}
+\bigg[im(\ln\frac{m^2}{\Lambda^2}+\gamma)d^{\mu}p
\nonumber\\
&&\hspace{0.5cm}
-2im(\ln\frac{m^2}{\Lambda^2}+\gamma+1)(a^{\mu}s+sa^{\mu})
-\frac{i}{3}(\ln\frac{m^2}{\Lambda^2}+\gamma+1)
d^{\nu}(d_{\nu}a^{\mu}-d_{\mu}a^{\nu})
\nonumber\\
&&\hspace{0.5cm}
+\frac{1}{3}(\ln\frac{m^2}{\Lambda^2}+\gamma+2)
(a^{\nu}V_{\mu\nu}-V_{\mu\nu}a^{\nu})
-\frac{i}{3}d^{\mu}d^{\nu}a_{\nu}
+\frac{i}{3}(\ln\frac{m^2}{\Lambda^2}+\gamma+4)(a^2a^{\mu}+a^{\mu}a^2)
\nonumber\\
&&\hspace{0.5cm}
-\frac{2i}{3}(\ln\frac{m^2}{\Lambda^2}+\gamma+2)a_{\nu}a^{\mu}a^{\nu}
-\epsilon_{\mu}^{~\sigma\rho\nu}
[-\frac{i}{4}(\ln\frac{m^2}{\Lambda^2}+\gamma)d^{\nu}V_{\sigma\rho}
-\frac{1}{3}a_{\sigma}d^{\nu}a_{\rho}+\frac{1}{3}(d_{\sigma}a_{\rho})a^{\nu}]
\bigg]\gamma_5\gamma^{\mu}\bigg\}\nonumber\\
&&\hspace{0.5cm}
+\mbox{tensor terms}+O(p^4)
\end{eqnarray}
One can easily check that above result can exactly reproduce
results (\ref{scalar}), (\ref{pseudo}),(\ref{vector}) and
(\ref{axial}). So two kinds calculation give same results for
coincidence limit of fermion propagator.

\section{Generalized  Schwinger proper time formulation
for Dirac operator with  dynamical fermion self energy}

In quantum field theory (\ref{path})
without bare fermion mass, the Lagrangian is chiral
symmetric. Usually the realization of chiral symmetry of the system is
characterized by symmetry order parameter
$\langle0|\overline{\psi}\psi|0\rangle$. If
$\langle0|\overline{\psi}\psi|0\rangle\neq 0$, we say the chiral symmetry is dynamically broken. This happens only when fermion has a nonzero self energy
$\Sigma(k^2)$ which is in general real and momentum dependent. The value of
$\Sigma(k^2)$ can be determined through solving Schwinger-Dyson equation.
If we need to use a phenomenological Lagrangian
to represent this DCSB effect, since momentum dependent fermion self energy
$\Sigma(k^2)$ in coordinate space is represented by
$\Sigma(-\partial_x^2)\delta(x-y)$, the naive generalization of fermion
interaction for explicit chiral symmetry breaking Lagrangian
$\overline{\psi}(D+m)\psi$ to DCSB Lagrangian should be
$\overline{\psi}[D+\Sigma(-\partial^2)]\psi$.
We argue this is not suitable since original
 $\overline{\psi}(D+m)\psi$ is invariant under $U_V(N_f)$ local symmetry
 transformation,
 \begin{eqnarray}
 \psi(x)&\rightarrow& \psi'(x)=h^{\dagger}(x)\psi(x)\nonumber\\
 s(x)&\rightarrow& s'(x)=h^{\dagger}(x)s(x)h(x)\nonumber\\
 p(x)&\rightarrow& p'(x)=h^{\dagger}(x)p(x)h(x)\label{vtrans}\\
 v_{\mu}(x)&\rightarrow& v_{\mu}'(x)=h^{\dagger}(x)v_{\mu}(x)h(x)
 +h ^{\dagger}(x)[i\partial_{\mu}h(x)]\nonumber\\
 a_{\mu}(x)&\rightarrow& a_{\mu}'(x)=h^{\dagger}(x)a_{\mu}(x)h(x)\;\;,\nonumber
  \end{eqnarray}
which leads to
\begin{eqnarray}
D_x+m\rightarrow (D_x+m)'&\equiv&
\partial\!\!\!/\;_x-iv\!\!\! /\;'(x)-ia\!\!\! /\;'(x)\gamma_5+m
-s'(x)+ip'(x)\gamma_5\nonumber\\
&=&h^{\dagger}(x)[\partial\!\!\!/\;_x-iv\!\!\! /\;(x)-ia\!\!\! /\;(x)\gamma_5
-s(x)+ip(x)\gamma_5+m]h(x)\nonumber\\
&=&h^{\dagger}(x)(D_x+m)h(x)\label{Dplusm}
\end{eqnarray}
Here the covariance is due to the property of constant $m$ which leads
\begin{eqnarray}
h^{\dagger}(x)mh(x)=m\;\;,
\end{eqnarray}
and then $\overline{\psi}(D+m)\psi$ is invariant
\begin{eqnarray}
\overline{\psi}(D+m)\psi\rightarrow
\overline{\psi}'(D+m)'\psi'\equiv&
\overline{\psi}hh^{\dagger}(D+m)hh^{\dagger}\psi=
\overline{\psi}(D+m)\psi
\end{eqnarray}
If we change $m$ to $\Sigma(-\partial^2)$, relation (\ref{Dplusm}) is no
longer valid due to differential operator dependence of $\Sigma$,
\begin{eqnarray}
h^{\dagger}(x)\Sigma(-\partial^2_x)h(x)&=&
\Sigma[-h^{\dagger}(x)\partial^2_xh(x)]
=\Sigma\bigg[-\bigg(\partial_{\mu}+h^{\dagger}(x)[\partial_{\mu}h(x)]\bigg)^2
\bigg]\neq\Sigma(-\partial^2_x)\;\;.
\end{eqnarray}
So, naive generalization to DCSB Lagrangian
$\overline{\psi}[D+\Sigma(-\partial^2)]\psi$ is not invariant on $U_V(N_f)$
symmetry. To implement this local symmetry (\ref{vtrans}),
instead of considering
$\Sigma(-\partial^2)$, we need to consider $\Sigma(-\overline{\nabla}^2)$
in which
\begin{eqnarray}
\overline{\nabla}^{\mu}\equiv\partial^{\mu}-iv^{\mu}(x)\;\;,
\end{eqnarray}
the bar over $\nabla_{\mu}$ is to specify the difference of present derivative with that introduced in (\ref{nabladef}).
Use (\ref{vtrans}), we find $\overline{\nabla}_{\mu}$  transform
as
\begin{eqnarray}
\overline{\nabla}^{\mu}_x\rightarrow\overline{\nabla}^{\mu\prime}_x
\equiv\partial^{\mu}_x-iv^{\mu\prime}(x)
=h^{\dagger}(x)\overline{\nabla}^{\mu}_xh(x)\;\;.
\end{eqnarray}
Then
\begin{eqnarray}
\Sigma(-\overline{\nabla}^2_x)&\rightarrow&
\Sigma(-\overline{\nabla}^{2\prime}_x)=
\Sigma[-h^{\dagger}(x)\overline{\nabla}^2_xh(x)]=
h^{\dagger}(x)\Sigma(-\overline{\nabla}^2_x)h(x)
\end{eqnarray}
and
\begin{eqnarray}
 &&[D_x+ \Sigma(-\overline{\nabla}_x^2) ]
 \rightarrow [D_x'+ \Sigma(-\overline{\nabla}_x^{2\prime})]
   =  h^{\dagger}(x)[D_x+ \Sigma(-\overline{\nabla}_x^2)   ]h(x)\;\;.
\end{eqnarray}
So, $\overline{\psi}[D+ \Sigma(-\overline{\nabla}^2)]\psi$ is invariant under
transformation (\ref{vtrans}) and can be treated as correct
generalization of $\overline{\psi}(D+m)\psi$.
Note , in principle, one can add in $\overline{\nabla}^{\mu}$ other terms
which
are covariant under transformation (\ref{vtrans}), such as  $a_{\mu}\gamma_5$
multiply by a constant. For these kind terms, symmetry itself is not enough to
 fix them completely, therefore they can be treated
  as some extra interactions which are beyond our choice of
 $\overline{\nabla}^{\mu}$ and  $\Sigma(-\overline{\nabla}^2)$. Our choice of
 $\Sigma(-\overline{\nabla}^2)$ is the minimal generalization required by
  symmetry to incorporating in fermion self energy into the theory.

The generalized real part of fermion determinant now is
\begin{eqnarray}
{\rm Re}\ln{\rm Det}[D+\Sigma(-\overline{\nabla}^2)]
&=&\frac{1}{2}{\rm Tr}\ln\bigg[[D^{\dagger}+\Sigma(-\overline{\nabla}^2)]
[D+\Sigma(-\overline{\nabla}^2)]\bigg]\nonumber\\
&=&-\frac{1}{2}\lim_{\Lambda\rightarrow\infty}
\int_{\frac{1}{\Lambda^2}}^{\infty}\frac{d\tau}{\tau}~
{\rm Tr}e^{-\tau[\overline{E}
-\nabla^2+\Sigma^2(-\overline{\nabla}^2)
+Jg(\overline{\nabla}^2)+\tilde{g}(\overline{\nabla}^2)K
-d\!\!\! /\;\Sigma(-\overline{\nabla}^2)]}\label{lnSigma}
\end{eqnarray}
where
\begin{eqnarray}
\overline{E}-\nabla^2
+\Sigma^2(-\overline{\nabla}^2)
+Jg(\overline{\nabla}^2)+\tilde{g}(\overline{\nabla}^2)K
-d\!\!\! /\;\Sigma(-\overline{\nabla}^2)
=[D^{\dagger}+\Sigma(-\overline{\nabla}^2)][D+\Sigma(-\overline{\nabla}^2)]
\end{eqnarray}
and
\begin{eqnarray}
&&[d\!\!\! /\;\Sigma(-\overline{\nabla}^2)]\equiv
\gamma^{\mu}[d_{\mu}\Sigma(-\overline{\nabla}^2)]=
\gamma^{\mu}\bigg(\partial_{\mu}\Sigma(-\overline{\nabla}^2)
-i[v_{\mu},\Sigma(-\overline{\nabla}^2)]\bigg)\nonumber\\
&&\overline{E}\equiv\frac{i}{4}[\gamma^{\mu},\gamma^{\nu}] {\cal
R}_{\mu\nu}+\gamma_{\mu}d^{\mu}(s-ip\gamma_5)
+i\gamma^{\mu}[a_{\mu}\gamma_5(s-ip\gamma_5)+(s-ip\gamma_5)a_{\mu}\gamma_5]
+s^2+p^2-[s,p]i\gamma_5\nonumber\\
&&g(x)=\tilde{g}(x)\equiv\Sigma(-x)\hspace{1.5cm} J=-ia\!\!\!
/\;\gamma_5-s-ip\gamma_5\hspace{2cm} K=-ia\!\!\!
/\;\gamma_5-s+ip\gamma_5\;\;.\nonumber
\end{eqnarray}
The generalized coincidence limit of fermion propagator is
\begin{eqnarray}
&&[D+\Sigma(-\overline{\nabla}^2)]^{-1}(x,x)\nonumber\\
&&=\bigg[\bigg([D^{\dagger}+\Sigma(-\overline{\nabla}^2)]
[D+\Sigma(-\overline{\nabla}^2)]\bigg)^{-1}
[D^{\dagger}+\Sigma(-\overline{\nabla}^2)]\bigg](x,x)\nonumber\\
&&=\lim_{\Lambda\rightarrow\infty}
\int_{\frac{1}{\Lambda^2}}^{\infty}d\tau~
\langle x|e^{-\tau[\overline{E}
-\nabla^2+\Sigma^2(-\overline{\nabla}^2)
+Jg(\overline{\nabla}^2)+\tilde{g}(\overline{\nabla}^2)K
-d\!\!\! /\;\Sigma(-\overline{\nabla}^2)]}
[-\nabla\!\!\!\! /\;-s(x)-ip(x)\gamma_5\nonumber\\
&&\hspace{0.5cm}
-2ia\!\!\! /\;(x)\gamma_5
+\Sigma(-\overline{\nabla}^2)]|x\rangle\label{propagenexp}
\end{eqnarray}
For safety of further calculation, we limit ultraviolet behavior of
$\Sigma(k^2)$ satisfy constraint
\begin{eqnarray}
\frac{\Sigma^2(k^2)}{k^2}\stackrel{k^2\rightarrow\infty}{--\rightarrow}0
\label{asymp}
\end{eqnarray}
For fermion determinant and coincidence limit of propagator,
 (\ref{lnSigma}) and (\ref{propagenexp}) tell us that the key now is to
 calculate matrix element of
\begin{eqnarray}
&&\langle x|e^{-\tau[\overline{E}
-\nabla^2+\Sigma^2(-\overline{\nabla}^2)
+Jg(\overline{\nabla}^2)+\tilde{g}(\overline{\nabla}^2)K
-d\!\!\! /\;\Sigma(-\overline{\nabla}^2)]}|x\rangle\nonumber\\
&&{\rm and}\label{inv}\\
&&\langle x|e^{-\tau[\overline{E}
-\nabla^2+\Sigma^2(-\overline{\nabla}^2)
+Jg(\overline{\nabla}^2)+\tilde{g}(\overline{\nabla}^2)K
-d\!\!\! /\;\Sigma(-\overline{\nabla}^2)]}
[-\nabla\!\!\!\! /\;
-s(x)-ip(x)\gamma_5-2ia\!\!\! /\;(x)\gamma_5
+\Sigma(-\overline{\nabla}^2)]|x\rangle\nonumber
\end{eqnarray}
which are much  more complex than (\ref{Eexp}) and (\ref{E1exp}). Since
now the operator on the exponential is beyond original
second- order elliptic partial differential operator, $\Sigma$ in
principle include arbitrary high order of differential operator
and except the constraint (\ref{asymp}), detail function $\Sigma$
on the differential operator is still unspecified.

In original Schwinger proper time formulation, there are two methods
 to calculate matrix elements (\ref{Eexp}) and (\ref{E1exp}).
 One is based on recursion formula and the other is
 momentum space calculation. Recursion formula can be obtained only if we can
obtain the matrix element with vanishing
 external fields which now is impossible, since
unspecified detail momentum dependence of $\Sigma(k^2)$ prevent us to obtain
it. In this work, we take second method to calculate
the matrix element in momentum space,
\begin{eqnarray}
&&\langle x|e^{-\tau[\overline{E}
-\nabla^2+\Sigma^2(-\overline{\nabla}^2)
+Jg(\overline{\nabla}^2)+\tilde{g}(\overline{\nabla}^2)K
-d\!\!\! /\;\Sigma(-\overline{\nabla}^2)]}|x\rangle\nonumber\\
&&=\int\frac{d^4k}{(2\pi)^4}\exp\bigg\{-\tau\bigg[\overline{E}(x)
-\nabla_x^2-2ik\cdot\nabla_x+k^2
+\Sigma^2(-\overline{\nabla}^2-2ik\cdot\overline{\nabla}_x+k^2)
\nonumber\\
&&\hspace{0.5cm}
+Jg(\overline{\nabla}_x^2+2ik\cdot\overline{\nabla}_x-k^2)
+\tilde{g}(\overline{\nabla}_x^2+2ik\cdot\overline{\nabla}_x-k^2)K
-d\!\!\! /\;\Sigma(-\overline{\nabla}_x^2-2ik\cdot\overline{\nabla}_x+k^2)
\bigg]\bigg\}\;\;.\label{momexp}
\end{eqnarray}
Assigning $\overline{E},J,K$ to be order of $p$, we can
take low energy expansion for (\ref{momexp}).
The result is given in appendix \ref{Seelyexp}. Similarly
\begin{eqnarray}
&&\langle x|e^{-\tau[\overline{E}
-\nabla^2+\Sigma^2(-\overline{\nabla}^2)
+Jg(\overline{\nabla}^2)+\tilde{g}(\overline{\nabla}^2)K
-d\!\!\! /\;\Sigma(-\overline{\nabla}^2)]}
[-\nabla\!\!\!\! /\;
-s(x)-ip(x)\gamma_5-2ia\!\!\! /\;(x)\gamma_5
+\Sigma(-\overline{\nabla}^2)]|x\rangle\nonumber\\
&&=\int\frac{d^4k}{(2\pi)^4}\exp\bigg\{-\tau\bigg[\overline{E}(x)
-\nabla_x^2-2ik\cdot\nabla_x+k^2
+\Sigma^2(-\overline{\nabla}^2-2ik\cdot\overline{\nabla}_x+k^2)
\nonumber\\
&&\hspace{0.5cm}
+Jg(\overline{\nabla}_x^2+2ik\cdot\overline{\nabla}_x-k^2)
+\tilde{g}(\overline{\nabla}_x^2+2ik\cdot\overline{\nabla}_x-k^2)K
-d\!\!\! /\;\Sigma(-\overline{\nabla}_x^2-2ik\cdot\overline{\nabla}_x+k^2)
\bigg]\bigg\}
\nonumber\\
&&\hspace{0.5cm}
\times[-\nabla\!\!\!\! /\;_x-ik\!\!\! /\;
-s(x)-ip(x)\gamma_5-2ia\!\!\! /\;(x)\gamma_5
+\Sigma(-\overline{\nabla}^2_x
-2ik\cdot\overline{\nabla}_x+k^2)]
|x\rangle\;\;.\label{momexp1}
\end{eqnarray}
We donot write down the result of expansion for this matrix element
explicitly. Instead we give  the final result
in terms of external fields $s,p,v_{\mu},a_{\mu}$
for coincidence limit of fermion propagator and discuss it in
section V.

We must note that the difficulty of present calculation is that the
momentum integration now can not be finished due to unknown behavior of
$\Sigma(k^2)$. While the chiral covariance of the result rely on the
achievement of momentum integration in original Schwinger proper time
formulation. We need to find a way to keep the covariance of chiral symmetry
 before finishing the momentum integration. Fortunately,
we found that those non-covariant
terms are all reduced to total divergence terms at momentum space which with
constraint (\ref{asymp}) vanish. So with invention of these total divergence
terms, we can still obtain a chiral covariant result as that in conventional
Schwinger proper time method even before we analytically finish the momentum
integration.

\section{Fermion determinant with momentum dependent fermion self energy}

We can parametrize the result fermion determinant as
\begin{eqnarray}
&&{\rm Re}\ln{\rm Det}[D+\Sigma(-\overline{\nabla}^2)]\nonumber\\
&&=\int d^4x{\rm  tr}_f\bigg[{\cal C}_0s+{\cal C}_1a^2+
 {\cal C}_2[d_{\mu}a^{\mu}]^2
 +{\cal C}_3(d^{\mu}a^{\nu}-d^{\nu}a^{\mu})(d_{\mu}a_{\nu}-d_{\nu}a_{\mu})
+{\cal C}_4a^4\label{detresult}\\ &&\hspace{0.5cm}
 +{\cal C}_5a^{\mu}a^{\nu}a_{\mu}a_{\nu} +{\cal C}_6s^2+{\cal C}_7p^2
+{\cal C}_8sa^2 +{\cal C}_9V^{\mu\nu}V_{\mu\nu}+{\cal
C}_{10}V^{\mu\nu}a_{\mu}a_{\nu} +{\cal
C}_{11}pd_{\mu}a^{\mu}\bigg]+O(p^6)\nonumber
\end{eqnarray}
Substitute the definition of $\overline{E}$, $J$, $K$, $g$ and
$\tilde{g}$ into result (\ref{matrixexp}) given in appendix
\ref{Seelyexp}, we obtain  coefficients ${\cal C}_i$ which are
related to $\Sigma$ by
\begin{eqnarray}
{\cal C}_0&=&-4\int d\tilde{k} \Sigma_kX_k\nonumber\\
 {\cal C}_1&=&2\int d\tilde{k}\bigg[
(-2\Sigma^2_k+k^2\Sigma_k\Sigma'_k)X_k^2+(-2\Sigma^2_k+k^2\Sigma_k\Sigma'_k)
\frac{X_k}{\Lambda^2}\bigg]\nonumber\\
 {\cal C}_2&=&2\int d\tilde{k}\bigg[2A_kX_k^3
+2A_k\frac{X_k^2}{\Lambda^2}+A_k\frac{X_k}{\Lambda^4}
+\frac{k^2}{2}\Sigma'^2_k\frac{X_k}{\Lambda^2}+\frac{k^2}{2}\Sigma'^2_kX_k^2
\bigg]\nonumber\\
 {\cal C}_3&=&\int d\tilde{k}\bigg[2B_kX_k^3
 +2B_k\frac{X_k^2}{\Lambda^2}+B_k\frac{X_k}{\Lambda^4}
+\frac{k^2}{2}\Sigma^{\prime 2}_k\frac{X_k}{\Lambda^2}
+\frac{k^2}{2}\Sigma^{\prime 2}_kX_k^2\bigg]\nonumber\\
 {\cal C}_4&=&-2\int d\tilde{k}\bigg[
(\frac{4\Sigma^4_k}{3}+\frac{2k^2\Sigma^2_k}{3}+\frac{k^4}{18})(
 6X_k^4+\frac{6X_k^3}{\Lambda^2}+\frac{3X_k^2}{\Lambda^4}
+\frac{X_k}{\Lambda^6})-(4\Sigma^2_k+\frac{k^2}{2})(
2X_k^3\nonumber\\
&&+\frac{2X_k^2}{\Lambda^2}+\frac{X_k}{\Lambda^4})
+\frac{X_k}{\Lambda^2}+X_k^2\bigg]\nonumber\\
 {\cal C}_5&=&-\int d\tilde{k}\bigg[
(\frac{-4\Sigma^4_k}{3}-\frac{2k^2\Sigma^2_k}{3}+\frac{k^4}{18})(
6X_k^4+\frac{6X_k^3}{\Lambda^2}+\frac{3X_k^2}{\Lambda^4}
+\frac{X_k}{\Lambda^6})
+4\Sigma^2_k(2X_k^3+\frac{2X_k^2}{\Lambda^2}\nonumber\\
&&+\frac{X_k}{\Lambda^4})-\frac{X_k}{\Lambda^2}-X_k^2\bigg]\nonumber\\
{\cal C}_6&=&-2\int
d\tilde{k}\bigg[(3\Sigma^2_k-2k^2\Sigma_k\Sigma'_k)X_k^2
+[2\Sigma^2_k-k^2(1+2\Sigma_k\Sigma'_k)]\frac{X_k}{\Lambda^2}\bigg]\nonumber\\
{\cal C}_7&=&-2\int
d\tilde{k}\bigg[(\Sigma^2_k-2k^2\Sigma_k\Sigma'_k)X_k^2
-k^2(1+2\Sigma_k\Sigma'_k)\frac{X_k}{\Lambda^2}\bigg]\nonumber\\
{\cal C}_8&=&-4\int d\tilde{k}\bigg[(4\Sigma^3_k+k^2\Sigma_k)X_k^3
+(4\Sigma^3_k+k^2\Sigma_k)\frac{X_k^2}{\Lambda^2}
+(2\Sigma^3_k+\frac{1}{2}k^2\Sigma_k)\frac{X_k}{\Lambda^4}
-3\Sigma_k\frac{X_k}{\Lambda^2}\nonumber\\
&&-3\Sigma_kX_k^2\bigg]\nonumber\\
{\cal C}_9&=&\int
d\tilde{k}\bigg[
(\frac{1}{3}k^2\Sigma'_k\Sigma''_k-\frac{1}{3}k^2\Sigma_k\Sigma''_k)X_k
+(-C_k+D_k)\frac{X_k}{\Lambda^2}
-(C_k-D_k)X_k^2+2E_kX_k^3\nonumber\\
&&+2E_k\frac{X_k^2}{\Lambda^2}
+E_k\frac{X_k}{\Lambda^4}\bigg]\nonumber\\
 {\cal C}_{10}&=&4i\int d\tilde{k}\bigg[
2F_kX_k^3+2F_k\frac{X_k^2}{\Lambda^2}
+F_k\frac{X_k}{\Lambda^4}
+\frac{k^2}{2}\Sigma_k^{\prime 2}\frac{X_k}{\Lambda^2}
+\frac{k^2}{2}\Sigma^{\prime 2}_kX_k^2\bigg]
 \nonumber\\
{\cal C}_{11}&=&4\int d\tilde{k}\bigg[
(\Sigma_k-\frac{1}{2}k^2\Sigma'_k)\frac{X_k}{\Lambda^2}
+(\Sigma_k-\frac{1}{2}k^2\Sigma'_k)X_k^2\bigg]\label{Kresult}
\end{eqnarray}
where
\begin{eqnarray}
\int d\tilde{k}&\equiv&
\int\frac{d^4k}{(2\pi)^4}e^{-\frac{k^2+\Sigma^2(k^2)}{\Lambda^2}}
\label{measure}\\
\Sigma_k&\equiv&\Sigma(k^2)\hspace{2cm}
X_k\equiv\frac{1}{k^2+\Sigma^2(k^2)}
\end{eqnarray}
and $A_k,B_k,C_k,D_k,E_k,F_k$ depending on $\Sigma(k^2)$
are given in appendix \ref{Coef}.
We see that asymptotic behavior of $\Sigma(k^2)$ (\ref{asymp}) insure
factor $e^{-\frac{k^2+\Sigma^2(k^2)}{\Lambda^2}}$ appeared in integration
 measure (\ref{measure}) is an ultraviolet damping factor which will keep
 our momentum integration convergent.

(\ref{detresult}) and (\ref{Kresult}) are our final result for the real part of
fermion determinant with presence of dynamical quark self energy. The result in
this paper is only up to order of $p^4$, one can easily generalize the
calculation to higher orders of the momentum expansion. As a self check of theory, take $\Sigma(k^2)$ be constant $m$, in the limit of
$\Lambda^2\rightarrow\infty$, the momentum integration in (\ref{Kresult})
can be finished, the result gives
\begin{eqnarray}
&&{\cal C}_0\stackrel{\Sigma=m}{--\rightarrow}
-\frac{N_c}{4\pi^2}m[\Lambda^2+m^2(ln\frac{m^2}{\Lambda^2}+\gamma-1)]
\nonumber\\ &&{\cal C}_1\stackrel{\Sigma=m}{--\rightarrow}
\frac{N_c}{4\pi^2}m^2(ln\frac{m^2}{\Lambda^2}+\gamma)\nonumber\\
&&{\cal C}_2 \stackrel{\Sigma=m}{--\rightarrow}
\frac{N_c}{24\pi^2}\nonumber\\ && {\cal
C}_3\stackrel{\Sigma=m}{--\rightarrow}
\frac{N_c}{48\pi^2}(ln\frac{m^2}{\Lambda^2}+\gamma+1)\nonumber\\
&&{\cal C}_4\stackrel{\Sigma=m}{--\rightarrow}
\frac{N_c}{24\pi^2}(ln\frac{m^2}{\Lambda^2}+\gamma+4)\nonumber\\
&&{\cal C}_5\stackrel{\Sigma=m}{--\rightarrow}
-\frac{N_c}{24\pi^2}(ln\frac{m^2}{\Lambda^2}+\gamma+2)\nonumber\\
&&{\cal C}_6\stackrel{\Sigma=m}{--\rightarrow}
\frac{N_c}{8\pi^2}\bigg[\Lambda^2+m^2(3ln\frac{m^2}{\Lambda^2}+3\gamma-1)
\bigg]\nonumber\\ &&{\cal C}_7\stackrel{\Sigma=m}{--\rightarrow}
\frac{N_c}{8\pi^2}\bigg[\Lambda^2+m^2(ln\frac{m^2}{\Lambda^2}+\gamma-1)\bigg]
\nonumber\\ &&{\cal C}_8\stackrel{\Sigma=m}{--\rightarrow}
-\frac{N_c}{2\pi^2}m(ln\frac{m^2} {\Lambda^2}+\gamma+1)
\nonumber\\ &&{\cal C}_9\stackrel{\Sigma=m}{--\rightarrow}
\frac{N_c}{48\pi^2}(ln\frac{m^2} {\Lambda^2}+\gamma)\nonumber\\
&&{\cal C}_{10}\stackrel{\Sigma=m}{--\rightarrow}
-\frac{iN_c}{12\pi^2}(ln\frac{m^2}{\Lambda^2}+\gamma+2)\nonumber\\
&&{\cal C}_{11}\stackrel{\Sigma=m}{--\rightarrow}
-\frac{N_c}{4\pi^2}m(ln\frac{m^2}{\Lambda^2}+\gamma)\;\;.
\end{eqnarray}
Substitute them back into (\ref{detresult}),  we reproduce
original result (\ref{exp}). So our generalized formulation can easily recover the conventional Schwinger proper time result.

\section{Coincidence limit of Fermion propagator with momentum dependent fermion self energy}

The computation of (\ref{propagenexp}) shows that the scalar,
pseudoscalar and axial vector parts of  coincidence limit of fermion propagator are
\begin{eqnarray}
&&\frac{1}{2}(1)^{\xi\zeta}\bigg[
[D+\Sigma(-\overline{\nabla}^2)]^{-1,(b\zeta)(a\xi)}(x,x)
+[D^{\dagger}+\Sigma(-\overline{\nabla}^{\dagger 2})]^{-1,(b\zeta)(a\xi)}(x,x)
\bigg]\nonumber\\
&&=-\bigg[{\cal C}_0+2{\cal C}_6s+
{\cal C}_8a^2\bigg]^{ba}+O(p^4)\;\;,\\
&&\frac{1}{2}(\gamma_5)^{\xi\zeta}
\bigg[
[D+\Sigma(-\overline{\nabla}^2)]^{-1,(b\zeta)(a\xi)}(x,x)
-[D^{\dagger}+\Sigma(-\overline{\nabla}^{\dagger 2})]^{-1,(b\zeta)(a\xi)}(x,x)
\bigg]\nonumber\\
&&=-i\bigg[2{\cal C}_7p+{\cal C}_{11}
d^{\mu}a_{\mu}\bigg]^{ba}+O(p^4)\;\;,\\
&&\frac{1}{2}(\gamma_5\gamma_{\mu})^{\xi\zeta}\bigg[
[D+\Sigma(-\overline{\nabla}^2)]^{-1,(b\zeta)(a\xi)}(x,x)
+[D^{\dagger}+\Sigma(-\overline{\nabla}^{\dagger 2})]^{-1,(b\zeta)(a\xi)}(x,x)
\bigg]\nonumber\\
&&=-i\bigg[2{\cal C}_1a^{\mu}
-2{\cal C}_2[d^{\mu}d^{\nu}a_{\nu}]
+2{\cal C}_3d^{\nu}(d^{\mu}a_{\nu}-d_{\nu}a^{\mu})
+2{\cal C}_4(a^2a^{\mu}+a^{\mu}a^2)
+4{\cal C}_5a_{\nu}a^{\mu}a^{\nu}
\nonumber\\
&&\hspace{0.5cm}+{\cal C}_8(sa^{\mu}+a^{\mu}s)
+{\cal C}_{10}(a_{\nu}V^{\mu\nu}-V^{\mu\nu}a_{\nu})
-{\cal C}_{11}d^{\mu}p\bigg]^{ba}+O(p^5)\;\;,
\end{eqnarray}
where the coefficients ${\cal C}_i$ are given in last section. Note the
computation is done for fermion determinant and propagator independently, and
for scalar, pseudoscalar and axial vector parts, we recover the result of which we directly
performing functional differential with  scalar, pseudo scalar and axial
vector
fields for determinant. Unfortunately, this property no longer valid for vector
 part of propagator. We parametrize the corresponding
coincidence limit of fermion propagator as
\begin{eqnarray}
&&2(\gamma_{\mu})^{\xi\zeta}\bigg[
[D+\Sigma(-\overline{\nabla}^2)]^{-1,(b\zeta)(a\xi)}(x,x)
-[D^{\dagger}+\Sigma(-\overline{\nabla}^{\dagger 2})]^{-1,(b\zeta)(a\xi)}(x,x)
\bigg]\nonumber\\
&&=i\bigg[2i\overline{\cal C}_2[d_{\nu}a^{\nu},a^{\mu}]
+4i\overline{\cal C}_3[d^{\mu}a^{\nu}-d^{\nu}a^{\mu},a_{\nu}]
+4\overline{\cal C}_9d^{\nu}V^{\mu\nu}
+\overline{\cal C}_{10}d_{\nu}[a^{\mu},a^{\nu}]
+i\overline{\cal C}_{11}[p,a^{\mu}]\bigg]^{ba}\nonumber\\
&&\hspace{0.5cm}+O(p^5)\;\;,
\end{eqnarray}
The coefficients $\overline{\cal C}_i$ with overline on them are to specify
the difference with their original coefficient ${\cal C}_i$ from
functional differential with vector fields.  We list down
their values in appendix \ref{Cbar}

Now we discuss the reason to cause this difference.
With momentum dependent fermion self energy, the coincidence limit of fermion
 propagator can not be obtained directly as that in hard mass case
 from the functional differential of  external fields for fermion determinant.
Since the formula corresponding to (\ref{derhard}) now is changed to
\begin{eqnarray}
&&\frac{\delta {\rm Tr}\ln[D+\Sigma(-\overline{\nabla}^2)]}
{\delta J^{\sigma\rho}(x)}\nonumber\\
&&=\int d^4yd^4z~
[D+\Sigma(-\overline{\nabla}^2)]^{-1,\rho'\sigma'}(y,z)
\frac{\delta [D+\Sigma(-\overline{\nabla}^2)]^{\sigma'\rho'}(z,y)}
{\delta J^{\sigma\rho}(x)}\nonumber\\
&&=[D+\Sigma(-\overline{\nabla}^2)]^{-1,\rho\sigma}(x,x)
+\int d^4yd^4z~[D+\Sigma(-\overline{\nabla}^2)]^{-1,\rho'\sigma'}(y,z)
\frac{\delta [\Sigma(-\overline{\nabla}^2)]^{\sigma'\rho'}(z,y)}
{\delta J^{\sigma\rho}(x)}\label{sigmadep}
\end{eqnarray}
and its hermitian conjugate formula.
The second term in (\ref{sigmadep})
is a new term which rely on the momentum dependent
$\Sigma(k^2)$ and vanishes when self energy is independent of external fields.
Because of this extra
term,  functional differential of  external fields for fermion determinant no
longer give propagator. Note that $\Sigma(-\overline{\nabla}^2)$ in the second
term is independent of scalar, pseudo scalar and axial vector
fields and the corresponding differential with scalar, pseudo scalar and axial
 vector fields vanish,
so this extra term donot contribute to scalar, pseudo scalar and axial vector
parts of coincidence limit of fermion propagator. This explains why direct
calculation of scalar, pseudo scalar and axial vector parts of coincidence
limit of propagator can recover result of functional differential for
determinant.

Paramitrize the second term as
\begin{eqnarray}
&&\frac{1}{2}(\gamma_{\mu})^{\xi\zeta}
\int d^4yd^4z~\bigg[
[D+\Sigma(-\overline{\nabla}^2)]^{-1,\rho'\sigma'}(y,z)
\frac{\delta [\Sigma(-\overline{\nabla}^2)]^{\sigma'\rho'}(z,y)}
{\delta J^{(a\xi)(b\zeta)}(x)}\nonumber\\
&&-[D^{\dagger}+\Sigma(-\overline{\nabla}^{\dagger 2})]^{-1,\rho'\sigma'}(y,z)
\frac{\delta [\Sigma(-\overline{\nabla}^{\dagger 2})]^{\sigma'\rho'}(z,y)}
{\delta J^{\dagger(a\xi)(b\zeta)}(x)}\bigg]
\nonumber\\
&&=i\bigg[2i\tilde{\cal C}_2[d_{\nu}a^{\nu},a^{\mu}]
+4i\tilde{\cal C}_3[d^{\mu}a^{\nu}-d^{\nu}a^{\mu},a_{\nu}]
+4\tilde{\cal C}_9d^{\nu}V^{\mu\nu}
+\tilde{\cal C}_{10}d_{\nu}[a^{\mu},a^{\nu}]
+i\tilde{\cal C}_{11}[p,a^{\mu}]\bigg]^{ba}\nonumber\\
&&\hspace{0.5cm}+O(p^5)\;\;,\label{ctilde}
\end{eqnarray}
We can use our generalized Schwinger proper time method directly calculate these coefficients.
\begin{eqnarray}
&&{\rm Tr}\bigg[[D+\Sigma(-\overline{\nabla}^2)]^{-1}
\frac{\delta[D+\Sigma(-\overline{\nabla}^2)]}{\delta J^{\sigma\rho}(x)}\bigg]
\nonumber\\
&&={\rm Tr}\bigg[\bigg([D^{\dagger}+\Sigma(-\overline{\nabla}^2)]
[D+\Sigma(-\overline{\nabla}^2)]\bigg)^{-1}
[D^{\dagger}+\Sigma(-\overline{\nabla}^2)]
\frac{\delta[D+\Sigma(-\overline{\nabla}^2)]}{\delta J^{\sigma\rho}(x)}\bigg]
\nonumber\\
&&=\lim_{\Lambda\rightarrow\infty}
\int_{\frac{1}{\Lambda^2}}^{\infty}d\tau\int d^4y~{\rm tr}
\langle y|e^{-\tau[\overline{E}
-\nabla^2+\Sigma^2(-\overline{\nabla}^2)
+Jg(\overline{\nabla}^2)+\tilde{g}(\overline{\nabla}^2)K
-d\!\!\! /\;\Sigma(-\overline{\nabla}^2)]}
[-\nabla\!\!\!\! /\;-s(x)-ip(x)\gamma_5\nonumber\\
&&\hspace{0.5cm}-2ia\!\!\! /\;(x)\gamma_5
+\Sigma(-\overline{\nabla}^2)]
\frac{\delta[D+\Sigma(-\overline{\nabla}^2)]}{\delta J^{\sigma\rho}(x)}
|y\rangle\;\;.
\end{eqnarray}
Combine above result and its hermitian conjugate together, we can calculate
 coefficients $\tilde{\cal C}$ in (\ref{ctilde}).
We list down the result coefficients $\tilde{\cal C}$ in appendix \ref{Cbar}.
One can check that summation over $\overline{\cal C}$ and $\tilde{\cal C}$
coefficients
together recover original $C$ coefficients:
\begin{eqnarray}
{\cal C}_i=\overline{\cal C}_i+\tilde{\cal C}_i\hspace{2cm}i=2,3,9,10,11
\end{eqnarray}

\section{Summary and discussion}

In this paper, we have generalized conventional Schwinger proper
time method for standard Dirac operator to incorporate dynamical
chiral symmetry breaking. The physical output of this
generalization is the real part of logarithm of fermion
determinant and coincidence limit of fermion propagator in
presence of momentum dependent fermion self energy $\Sigma(k^2)$.
The mathematical progress of this generalization is that the
operator on the exponential in the key matrix element of Schwinger
proper time formulation is now generalized from original
second-order elliptic differential operator $E-\nabla^2$ (see
(\ref{Seely})) to arbitrary high order differential operator
$\overline{E}-\nabla^2+\Sigma^2(-\overline{\nabla}^2)
+Jg(\overline{\nabla}^2)+\tilde{g}(\overline{\nabla}^2)K -d\!\!\!
/\;\Sigma(-\overline{\nabla}^2)$ (see (\ref{inv})). This high
order differential  operator dependence is represented by an
unspecified function $\Sigma(-\overline{\nabla}^2)$ which
physically is fermion self energy and characterize the dynamical
chiral symmetry breaking of the system. The generalized formulation
automatically keep the local symmetry $U_V(N_f)$ given by
(\ref{vtrans}) which is residual symmetry after explicitly
breaking of local chiral symmetry $U_L(N_f)\otimes U_R(N_f)$ by
constant fermion mass.

As discussed in the introduction of this paper, one can further generalize this local symmetry $U_V(N_f)$ invariance of the formulation
 to the invariance of full chiral symmetry $U_L(N_f)\otimes U_R(N_f)$.
 The price is that the symmetry must be realized nonlinearly, i.e.,
 we need to introduce into theory a local field $\Omega(x)$
 which  transform nonlinearly  under local chiral symmetry (\ref{chiral}) as
 (\ref{Omega}) and rotated external fields defined as (\ref{JOmegadef}).
Replacing the external fields in $D+\Sigma(-\overline{\nabla}^2)$ in
(\ref{lnSigma}) and (\ref{propagenexp}) with
$D_{\Omega}+\Sigma(-\overline{\nabla}_{\Omega}^2)$, $
\overline{\nabla}^{\mu}_{\Omega}\equiv\partial^{\mu}-iv^{\mu}_{\Omega}(x)$.
One can easily check that under local  chiral symmetry
$U_L(N_f)\otimes U_R(N_f)$ transformation (\ref{chiral}) and (\ref{Omega}),
we have  (\ref{hidden}), (\ref{hidden1}) and
\begin{eqnarray}
D_{\Omega}+\Sigma(-\overline{\nabla}_{\Omega}^2)
&\rightarrow&
D_{\Omega}'+\Sigma(-\overline{\nabla}_{\Omega}^{2\prime})
=h^{\dagger}(x)
[D_{\Omega}+\Sigma(-\overline{\nabla}_{\Omega}^2)]h(x)\label{hidden2}
\end{eqnarray}
i.e., they are covariant quantities. In terms of rotated quantities,
 once our formulation keep this local hidden  symmetry $U_V(N_f)$, it
 automatically keep original  local  chiral symmetry
 $U_L(N_f)\otimes U_R(N_f)$.

In this paper, we do not calculate the imaginary part of fermion determinant and fermion propagator are only limited to its coincidence limit, its general nonlocal part is not computed.These are all under investigation and we will present results in separated papers.

\section*{Acknowledgments}

This work was  supported by the National Natural Science Foundation of China,
the Foundation of Fundamental Research Grant of Tsinghua University.



\appendix

\section{Coefficients definitions}\label{Coef}

\begin{eqnarray}
A_k&=&\frac{2}{3}k^2\Sigma_k\Sigma'_k(-1-2\Sigma_k\Sigma'_k)-\frac{1}{3}
\Sigma^2_k(-1-2\Sigma_k\Sigma'_k)
+\frac{1}{3}k^2\Sigma^2_k(\Sigma^{\prime 2}_k+
\Sigma_k\Sigma''_k)\nonumber\\
&&+\frac{1}{6}k^4(\Sigma^{\prime 2}_k+\Sigma_k\Sigma''_k)
\nonumber\\
B_k&=&\frac{2}{3}k^2\Sigma_k\Sigma'_k(-1-2\Sigma_k\Sigma'_k)-\frac{1}{3}
\Sigma^2_k(-1-2\Sigma_k\Sigma'_k)+\frac{1}{3}k^2\Sigma^2_k(\Sigma^{\prime 2}_k
+\Sigma_k\Sigma''_k)\nonumber\\
&&+\frac{1}{18}k^4(\Sigma^{\prime 2}_k+\Sigma_k\Sigma''_k)
+\frac{1}{6}k^2(-1-2\Sigma_k\Sigma'_k)\nonumber\\
C_k&=&\frac{1}{3}-\frac{1}{3}\Sigma_k\Sigma'_k
+\frac{1}{2}k^2\Sigma^{\prime 2}_k\nonumber\\
D_k&=&-\frac{1}{2}k^2\Sigma^{\prime 2}_k+\frac{1}{6}k^2\Sigma_k\Sigma''_k
(-1-2\Sigma_k\Sigma'_k)+\frac{2}{9}k^4\Sigma'_k\Sigma''_k
(1+2\Sigma_k\Sigma'_k)]
-\frac{2}{9}k^4\Sigma^{\prime 2}_k(-\Sigma^{\prime 2}_k-\Sigma_k\Sigma''_k)
\nonumber\\
&&+\frac{1}{3}k^2\Sigma_k\Sigma'_k(-\Sigma^{\prime 2}_k-\Sigma_k\Sigma''_k)
\nonumber\\
E_k&=&\frac{1}{6}k^2\Sigma_k\Sigma'_k(-1-2\Sigma_k\Sigma'_k)^2
-\frac{1}{9}k^4\Sigma^{\prime 2}_k(1+2\Sigma_k\Sigma'_k)^2\nonumber\\
F_k&=&\frac{4}{3}k^2\Sigma_k\Sigma'_k-
\frac{4}{3}k^2(\Sigma_k\Sigma'_k)^2-\frac{2}{3}\Sigma^2_k
+\frac{2}{3}\Sigma^3_k\Sigma'_k
+\frac{1}{3}k^2\Sigma^2_k(\Sigma^{\prime 2}_k+\Sigma_k\Sigma''_k)
+\frac{1}{9}k^4(\Sigma^{\prime 2}_k+\Sigma_k\Sigma''_k)\nonumber\\
&&+\frac{1}{3}k^2(-1-2\Sigma_k\Sigma'_k)+\frac{1}{2}k^2\nonumber
\end{eqnarray}

\section{Generalized Seely-DeWitt expansion}\label{Seelyexp}

The detail calculation gives:
\begin{eqnarray}
&&\langle x|e^{-\tau[\overline{E}-\nabla^2+\Sigma^2(-\overline{\nabla}^2)
+Jg(\overline{\nabla}^2)
+\tilde{g}(\overline{\nabla}^2)K-
d\!\!\! /\;\Sigma(-\overline{\nabla}^2)]}|x\rangle
\nonumber\\
&&=\int\frac{d^4k}{(2\pi)^4}e^{-\tau f}\bigg\{Y_1+k^2 Y_2+k^4
Y_3 +\frac{1}{8}k^2(-\frac{\tau}{3}f''+\tau^2f'f''-\frac{\tau^3}{3}f'^3)
[\overline{\nabla}^{\mu},\overline{\nabla}^{\nu}]
[\overline{\nabla}_{\mu},\overline{\nabla}_{\nu}]\nonumber\\
&&\hspace{0.5cm}-i\frac{\tau^2}{36}f'''k^4\gamma_5(
[\overline{\nabla}^{\mu},[\overline{\nabla}_{\mu},\overline{\nabla}^{\nu}]]
a_{\nu}+a_{\nu}[\overline{\nabla}^{\mu},[\overline{\nabla}_{\mu},
\overline{\nabla}^{\nu}]])+i\frac{\tau^3}{36}f'f''k^4\gamma_5(
-2[a^{\nu},[\overline{\nabla}^{\mu},[\overline{\nabla}_{\mu},
\overline{\nabla}_{\nu}]]]\nonumber\\
&&\hspace{0.5cm}+
[\overline{\nabla}^{\mu},\overline{\nabla}^{\nu}]
[\overline{\nabla}_{\nu},a_{\mu}]+[\overline{\nabla}_{\nu},a_{\mu}]
[\overline{\nabla}^{\mu},\overline{\nabla}^{\nu}]
+[\overline{\nabla}_{\nu},[\overline{\nabla}^{\mu},\overline{\nabla}^{\nu}]
a_{\mu}]+[\overline{\nabla}_{\nu},a_{\mu}[\overline{\nabla}^{\mu},
\overline{\nabla}^{\nu}]])
\nonumber\\
&&\hspace{0.5cm}-i\frac{\tau^4}{36}{f'}^3\gamma_5([
\overline{\nabla}_{\nu},a_{\mu}][\overline{\nabla}^{\mu},
\overline{\nabla}^{\nu}]+
[\overline{\nabla}^{\mu},\overline{\nabla}^{\nu}][\overline{\nabla}_{\nu},
a_{\mu}])-\frac{\tau^3}{12}{f'}^2k^2[\overline{\nabla}^{\mu},
[\overline{\nabla}_{\mu},a^2]])\nonumber\\
&&\hspace{0.5cm}
-\frac{\tau^3}{36}f''k^4([\overline{\nabla}^{\mu},a_{\mu}]
[\overline{\nabla}^{\nu},a_{\nu}]
+2[\overline{\nabla}^{\nu},a^{\mu}]
[\overline{\nabla}_{\mu},a_{\nu}]-2[\overline{\nabla}^{\mu},
[\overline{\nabla}^{\nu},a_{\mu}a_{\nu}]]
-2a_{\mu}a_{\nu}[\overline{\nabla}^{\mu},\overline{\nabla}^{\nu}]) \nonumber\\
&&\hspace{0.5cm}+
\frac{\tau^4}{36}f'^2k^4(a^{\nu}[\overline{\nabla}^{\mu},
[\overline{\nabla}_{\mu},a_{\nu}]]+[\overline{\nabla}^{\mu},
[\overline{\nabla}_{\mu},a_{\nu}]]a^{\nu}
+[\overline{\nabla}^{\mu},a^{\nu}][\overline{\nabla}_{\mu},a_{\nu}]
+[\overline{\nabla}^{\nu},a^{\mu}][\overline{\nabla}_{\mu},a_{\nu}]
\nonumber\\
&&\hspace{0.5cm}
-a_{\mu}a_{\nu}[\overline{\nabla}^{\mu},\overline{\nabla}^{\nu}]
-[\overline{\nabla}^{\mu},\overline{\nabla}^{\nu}]a_{\mu}a_{\nu})
-\frac{\tau^3}{12}f'k^2\gamma_5([a^{\mu},[\overline{\nabla}^{\mu},a^2]]
+[\overline{\nabla}^{\mu},[a_{\mu},a^2]])\nonumber\\
&&\hspace{0.5cm}
+\frac{i\tau^4}{36}f'k^4\gamma_5([\overline{\nabla}^{\mu},a^{\nu}]
[a_{\nu},a_{\mu}]+[a_{\nu},a_{\mu}][\overline{\nabla}^{\mu},a^{\nu}]
+[\overline{\nabla}^{\mu},[a_{\mu},a^2]]+[a^{\mu},[\overline{\nabla}^{\mu}
,a^2]])\nonumber\\
&&\hspace{0.5cm}
-\frac{\tau^3}{12}{f'}^2k^2[\overline{\nabla}^{\mu},
[\overline{\nabla}_{\mu},F]]
+\frac{\tau^2}{4}f'k^2[\overline{\nabla}^{\mu},[\overline{\nabla}_{\mu},F']]
+\frac{\tau^2}{4}k^2({g'}^2J[\overline{\nabla}^{\mu},
[\overline{\nabla}_{\mu},J]]\nonumber\\
&&\hspace{0.5cm}
+\tilde{g}'^2[\overline{\nabla}^{\mu},[\overline{\nabla}_{\mu},K]]K
+g'\tilde{g}'[\overline{\nabla}^{\mu},[\overline{\nabla}_{\mu},KJ]])
-\frac{\tau^3}{12}f'k^2(2F'[\overline{\nabla}^{\mu},
[\overline{\nabla}_{\mu},F]]
\nonumber\\
&&\hspace{0.5cm}
+2[\overline{\nabla}^{\mu},[\overline{\nabla}_{\mu},F']]F
+2[\overline{\nabla}^{\mu},F'][\overline{\nabla}_{\mu},F]
+2F[\overline{\nabla}^{\mu},[\overline{\nabla}_{\mu},F']]
+2[\overline{\nabla}^{\mu},[\overline{\nabla}_{\mu},F]]F'\nonumber\\
&&\hspace{0.5cm}
+2[\overline{\nabla}^{\mu},F][\overline{\nabla}_{\mu},F']
-[\overline{\nabla}^{\mu},[\overline{\nabla}_{\mu},Jg'F+F\tilde{g}'K]]
-[\overline{\nabla}^{\mu},[\overline{\nabla}_{\mu},F]]Jg'
-\tilde{g}'K[\overline{\nabla}^{\mu},[\overline{\nabla}_{\mu},F])
\nonumber\\
&&\hspace{0.5cm}
+\frac{\tau^4}{24}{f'}^2k^2(F[\overline{\nabla}^{\mu},
[\overline{\nabla}_{\mu},F]]+[\overline{\nabla}^{\mu},[\overline{\nabla}_{\mu},
F]]F+[\overline{\nabla}^{\mu},F][\overline{\nabla}_{\mu},F])
\nonumber\\
&&\hspace{0.5cm}
+\frac{\tau^2}{4}k^2\gamma_5[g'(a_{\mu}[\overline{\nabla}^{\mu},J]
+[\overline{\nabla}^{\mu},a_{\mu}J]-J[\overline{\nabla}^{\mu},a_{\mu}])
+\tilde{g}'([\overline{\nabla}^{\mu},a_{\mu}]K
-[\overline{\nabla}^{\mu},Ka_{\mu}]-[\overline{\nabla}^{\mu},K]a_{\mu})]
\nonumber\\
&&\hspace{0.5cm}
-\frac{i\tau^3}{12}f'k^2\gamma_5([\overline{\nabla}^{\mu},[a_{\mu},F]]
+[a_{\mu},[\overline{\nabla}^{\mu},F]])
-\frac{\tau^3}{12}k^2\gamma_5(F'[\overline{\nabla}^{\mu},[a_{\mu},F]]
 +[a_{\mu},[\overline{\nabla}^{\mu},F']]F \nonumber\\
&&\hspace{0.5cm}
+[a_{\mu},F'][\overline{\nabla}^{\mu},F]+F'[\overline{\nabla}^{\mu},
[a_{\mu},F]] +[\overline{\nabla}^{\mu},[a_{\mu},F']]F
+[\overline{\nabla}^{\mu},F'][a_{\mu},F]) \nonumber\\
&&\hspace{0.5cm}
-\frac{i\tau^3}{12}g'k^2\gamma_5 [g([\overline{\nabla}^{\mu},J]a_{\mu}F
 -a_{\mu}J[\overline{\nabla}^{\mu},F]-J[\overline{\nabla}^{\mu},a_{\mu}F]
 +[\overline{\nabla}^{\mu},J]Fa_{\mu}
+F[\overline{\nabla}^{\mu},J]a_{\mu}\nonumber\\
&&\hspace{0.5cm}
+[\overline{\nabla}^{\mu},Fa_{\mu}J]+a_{\mu}F[\overline{\nabla}^{\mu},J]
+[\overline{\nabla}^{\mu},FJ]a_{\mu})
+[\tilde{g}(-a_{\mu}[\overline{\nabla}^{\mu},KF]
-[\overline{\nabla}^{\mu},K]Fa_{\mu}
-[\overline{\nabla}^{\mu},Ka_{\mu}F]\nonumber\\
&&\hspace{0.5cm}
-a_{\mu}[\overline{\nabla}^{\mu},K]F -a_{\mu}F[\overline{\nabla}^{\mu},K]
+[\overline{\nabla}^{\mu},Fa_{\mu}]K
-Fa_{\mu}[\overline{\nabla}^{\mu},K]+[\overline{\nabla}^{\mu},F]Ka_{\mu})]
\nonumber\\
&&\hspace{0.5cm}
+\frac{i\tau^4}{24}f'k^2\gamma_5(F[\overline{\nabla}^{\mu},[a_{\mu},F]]
+[\overline{\nabla}^{\mu},[a_{\mu}\gamma_5,F]]F
+[\overline{\nabla}^{\mu},F][a_{\mu},F]+F[a_{\mu} ,[\overline{\nabla}^{\mu},F]]
\nonumber\\
&&\hspace{0.5cm}
+[a_{\mu},[\overline{\nabla}^{\mu},F]]F
+[a_{\mu},F][\overline{\nabla}^{\mu},F])
-\frac{\tau^2}{4}\Sigma'k^2(F'[\overline{\nabla}_{\mu},[-i\nabla\!\!\!\! /\;,
\overline{\nabla}^{\mu}]]-[\overline{\nabla}_{\mu}[-i\nabla\!\!\!\! /\;,
\overline{\nabla}^{\mu}]]F')\nonumber\\
&&\hspace{0.5cm}
+i\frac{\tau^2}{2}\Sigma'k^2((\tilde{g}[\overline{\nabla}^{\mu},K]
[\nabla\!\!\!\! /\;,\overline{\nabla}^{\mu}]+g[\nabla\!\!\!\! /\;,
\overline{\nabla}^{\mu}]][J,\overline{\nabla}^{\mu}])
+\frac{\tau^2}{2}\Sigma'k^2[\nabla\!\!\!\! /\;,\overline{\nabla}^\mu]^2
\nonumber\\
&&\hspace{0.5cm}
+i\frac{\tau^3}{12}f'\Sigma'
k^2([\overline{\nabla}_{\mu},[\nabla\!\!\!\! /\;,\overline{\nabla}^{\mu}]F]
+[\overline{\nabla}_{\mu},F[\nabla\!\!\!\! /\;,\overline{\nabla}^{\mu}]]
+[[\nabla\!\!\!\! /\;,\overline{\nabla}_{\mu}],[\overline{\nabla}^{\mu},F])
\nonumber\\
&&\hspace{0.5cm}
-\frac{\tau^2}{2}\Sigma'k^2\gamma_5([\nabla\!\!\!\! /\;,
\overline{\nabla}^{\mu}]a_{\mu}+a_{\mu}[\nabla\!\!\!\! /\;,
\overline{\nabla}^{\mu}])+\frac{\tau^3}{6}\Sigma'\gamma_5
k^2(a^{\mu}[\nabla\!\!\!\! /\;,\overline{\nabla}^{\mu}]F
+a^{\mu}F[\nabla\!\!\!\! /\;,\overline{\nabla}^{\mu}]\nonumber\\
&&\hspace{0.5cm}
+[\nabla\!\!\!\! /\;,\overline{\nabla}^{\mu}]Fa^{\mu}
+[\nabla\!\!\!\! /\;,\overline{\nabla}^{\mu}]a^{\mu}F
+Fa^{\mu}[\nabla\!\!\!\! /\;,\overline{\nabla}^{\mu}]
+F[\nabla\!\!\!\! /\;,\overline{\nabla}^{\mu}]a^{\mu})\bigg\}\nonumber\\
&&\hspace{0.5cm}+\mbox{total derivative terms}\label{matrixexp}
\end{eqnarray}
where the total derivative terms are
\begin{eqnarray}
&&\mbox{total derivative terms}\nonumber\\
&&=\int\frac{d^4k}{(2\pi)^4}\frac{\partial}{\partial k^{\mu}}k^{\mu}\bigg\{
e^{-\tau f}[\frac{\tau}{72}k^2f'''
+\frac{\tau^2}{8}f''-\frac{\tau^2}{24}k^2f'f''
+\frac{\tau^3}{72}k^2 f'^3
+\frac{\tau^2}{8} f'^2]\nabla^4_x \nonumber\\
&&\hspace{0.5cm}-k^2 e^{-\tau f}[-\frac{\tau}{72}f'''
+\frac{\tau^2}{24}f'f''-\frac{\tau^3}{72}f'^3]{\overline{\nabla}^{\mu}}
{\overline{\nabla}^2}{\overline{\nabla}_{\mu}}\nonumber\\
&&\hspace{0.5cm}
-k^2 e^{-\tau f}[-\frac{\tau}{72}f'''
+\frac{\tau^2}{24}f'f''-\frac{\tau^3}{72}f'^3]
{\overline{\nabla}^{\mu}}{\overline{\nabla}^{\nu}}{\overline{\nabla}_{\mu}}
{\overline{\nabla}_{\nu}}
-\frac{i\tau}{4}k^2\gamma_5 e^{-\tau f}
(a^{\mu}\overline{\nabla}_{\mu}+\overline{\nabla}^{\mu}a_{\mu})\nonumber\\
&&\hspace{0.5cm}+\frac{i\tau^2}{8}k^2\gamma_5 e^{-\tau f}f'
[\overline{\nabla}^2(a^{\mu}\overline{\nabla}_{\mu}+\overline{\nabla}^{\mu}
a_{\mu})
+(a^{\mu}\overline{\nabla}_{\mu}+\overline{\nabla}^{\mu}a_{\mu})
\overline{\nabla}^2]\nonumber\\
&&\hspace{0.5cm}
-\frac{i\tau^2\gamma_5}{24}k^2 e^{-\tau f}f''[
\overline{\nabla}_{\mu}\overline{\nabla}^2a^{\mu}
+\overline{\nabla}^2\overline{\nabla}^{\mu}a_{\mu}
+a^{\mu}\overline{\nabla}_{\mu}\overline{\nabla}^2+
a_{\mu}\overline{\nabla}^2\overline{\nabla}^{\mu}]\nonumber\\
&&\hspace{0.5cm}
+\frac{i\tau^3\gamma_5}{72}k^2 e^{-\tau f}f'^2
[\overline{\nabla}^2(a^{\mu}\overline{\nabla}_{\mu}+\overline{\nabla}^{\mu}
a_{\mu})+(a^{\mu}\overline{\nabla}_{\mu}+\overline{\nabla}^{\mu}
a_{\mu})\overline{\nabla}^2
+2\overline{\nabla}^{\mu}(a^{\mu}\overline{\nabla}_{\mu}
+\overline{\nabla}^{\mu}a_{\mu})\overline{\nabla}_{\mu}\nonumber\\
&&\hspace{0.5cm}
+2\overline{\nabla}_{\mu}\overline{\nabla}^2
+2a_{\mu}\overline{\nabla}^2\overline{\nabla}^{\mu}]
+\frac{\tau^2}{8}k^2 e^{-\tau f(-k^2)}f'(-k^2)
[\overline{\nabla}^2a^2+a^2\overline{\nabla}^2]\nonumber\\
&&\hspace{0.5cm}
+\frac{i\tau^2\gamma_5}{8} e^{-\tau f}[a^2(a^{\mu}\overline{\nabla}_{\mu}
+\overline{\nabla}^{\mu}a_{\mu})+(a^{\mu}\overline{\nabla}_{\mu}
+\overline{\nabla}^{\mu}a_{\mu})a^2]
-\frac{\tau^2}{8} e^{-\tau f}(\overline{\nabla}^{\mu}a_{\mu}
+a^{\mu}\overline{\nabla}_{\mu})^2\nonumber\\
&&\hspace{0.5cm}
-\frac{\tau^3}{72}k^2 e^{-\tau f}f'[(\overline{\nabla}^{\mu}
a_{\mu}+a^{\mu}\overline{\nabla}_{\mu})^2
+2\overline{\nabla}^{\nu}(\overline{\nabla}^{\mu}a_{\mu}
+a^{\mu}\overline{\nabla}_{\mu})a_{\nu}
+2a^{\nu}(\overline{\nabla}^{\mu}a_{\mu}+a^{\mu}\overline{\nabla}_{\mu})
\overline{\nabla}_{\nu}+2a^2\overline{\nabla}^2\nonumber\\
&&\hspace{0.5cm}
+2\overline{\nabla}^2a^2+2a^{\mu}\overline{\nabla}^2a_{\mu}]
-\frac{i\tau^3}{36} e^{-\tau f}[a^2(a^{\mu}\overline{\nabla}_{\mu}
+\overline{\nabla}^{\mu}a_{\mu})+(a^{\mu}\overline{\nabla}_{\mu}
+\overline{\nabla}^{\mu}a_{\mu})a^2+a^{\nu}(a^{\mu}\overline{\nabla}_{\mu}
+a^{\mu}a_{\mu})a_{\nu}]\nonumber\\
&&\hspace{0.5cm}
-\frac{\tau}{4} e^{-\tau f}(g'J\overline{\nabla}^2
+\tilde{g}\overline{\nabla}^2K)+\frac{\tau^2}{8} e^{-\tau f}f'
(F\overline{\nabla}^2+\overline{\nabla}^2F) \nonumber\\
&&\hspace{0.5cm}
+\frac{\tau^2}{8} e^{-\tau f}[g'(J\overline{\nabla}^2F
+FJ\overline{\nabla}^2)+\tilde{g}(\overline{\nabla}^2KF+F\overline{\nabla}^2K)]
-\frac{\tau^3}{24} e^{-\tau f}(\overline{\nabla}^2F^2
+F^2\overline{\nabla}^2+F\overline{\nabla}^2F)\nonumber\\
&&\hspace{0.5cm}
+\frac{i\tau^2\gamma_5}{8} e^{-\tau f}([a^{\mu}\overline{\nabla}_{\mu}
+\overline{\nabla}_{\mu}a^{\mu}]F+F[a^{\mu}\overline{\nabla}_{\mu}
+\overline{\nabla}_{\mu}a^{\mu}])\nonumber\\
&&\hspace{0.5cm}
-\frac{i\tau^3\gamma_5}{24} e^{-\tau f}(
[a^{\mu}\overline{\nabla}_{\mu}+\overline{\nabla}_{\mu}a^{\mu}]F^2
+F^2[a^{\mu}\overline{\nabla}_{\mu}+\overline{\nabla}_{\mu}a^{\mu}]
+F[a^{\mu}\overline{\nabla}_{\mu}+\overline{\nabla}_{\mu}a^{\mu}]F)
\nonumber\\
&&\hspace{0.5cm}
+i\frac{\tau}{4} e^{-\tau f}\Sigma'[\nabla\!\!\!\! /\;,
\overline{\nabla}^2]-i\frac{\tau^2}{8} e^{-\tau f}\Sigma'
([\nabla\!\!\!\! /\;,\overline{\nabla}^2]F+F[\nabla\!\!\!\! /\;,
\overline{\nabla^2}])\bigg\}
\end{eqnarray}
and
\begin{eqnarray}
&&f\equiv k^2+\Sigma^2(k^2)\hspace{1cm}
f'\equiv-\frac{\partial f}{\partial k^2}\hspace{1cm}
f''\equiv\frac{\partial^2 f}{\partial (k^2)^2}\hspace{1cm}
f^{\prime\prime\prime}\equiv-\frac{\partial^3 f}{\partial (k^2)^3}\nonumber\\
&&F=\overline{E}(x)+J(x)g(-k^2)+\tilde{g}(-k^2)K(x)\nonumber\\
&&F'=J(x)g'(-k^2)+\tilde{g}'(-k^2)K(x)\nonumber\\
&&Y_1(x)=1-\tau[a^2(x)+F(x)]+\frac{\tau^2}{2}[a^2(x)+F(x)]^2
-\frac{\tau^3}{6}[a^2(x)F^2(x)+F^2(x)a^2(x)\nonumber\\
&&\hspace{1.5cm}+F(x)a^2(x)F(x)+F^3(x)]
+\frac{\tau^4}{4!}F^4(x)\nonumber\\
&&Y_2(x)=\frac{\tau^2}{2}a^2(x)-\frac{\tau^3}{6}[2a^4(x)+a^{\mu}(x)
 a^2(x)a_{\mu}(x)+F(x)a^2(x)+a^2(x)F(x)+a^{\mu}F(x)a_{\mu}(x)]\nonumber\\
&&\hspace{1.5cm}-\frac{\tau}{4}[F^2(x)a^2(x)+a^2(x)F^2(x)
+a^{\mu}(x)F^2(x)a_{\mu}(x)
+F(x)a^2(x)F(x)\nonumber\\
&&\hspace{1.5cm}+F(x)a^{\mu}(x)F(x)a_{\mu}(x)+a^{\mu}(x)F(x)a_{\mu}(x)F(x)]
\nonumber\\
&&Y_3(x)=\frac{\tau^4}{4!}[a^4(x)+a^{\mu}(x)a^{\nu}(x)a_{\mu}(x)a_{\nu}(x)
+a^{\mu}(x)a^2(x)a_{\mu}(x)]\nonumber
\end{eqnarray}

\section{$\overline{\cal C}$ and $\tilde{\cal C}$  coefficients}\label{Cbar}

$\overline{\cal C}$ coefficients are defined as
\begin{eqnarray}
\overline{\cal C}_2&=&2\int d\tilde{k}~
\overline{A}_k
[2X_k^3+2\frac{X_k^2}{\Lambda^2}+\frac{X_k}{\Lambda^4}]\nonumber\\
\overline{\cal C}_3&=&\int d\tilde{k}~
[\overline{B}_k(2X_k^3+2\frac{X_k^2}{\Lambda^2}+\frac{X_k}{\Lambda^4})+X_k^2+
\frac{X_k}{\Lambda^2}]\nonumber\\
\overline{\cal C}_9&=&-\int d\tilde{k}~
\overline{C}_k[\frac{X_k}{\Lambda^2}+X_k^2]\nonumber\\
\overline{\cal C}_{10}&=&4i\int d\tilde{k}~
\overline{F}_k[2X_k^3+2\frac{X_k^2}{\Lambda^2}+\frac{X_k}{\Lambda^4}]
\nonumber\\
\overline{\cal C}_{11}&=&4\int d\tilde{k}~\Sigma_k[\frac{X_k}{\Lambda^2}+X_k^2]
\end{eqnarray}
with
\begin{eqnarray}
\overline{A}_k&=&\frac{1}{6}[2\Sigma_k^2-k^2\Sigma_k\Sigma'_k]\nonumber\\
\overline{B}_k&=&\frac{1}{3}[k^2+\Sigma_k^2+k^2\Sigma_k\Sigma'_k]\nonumber\\
\overline{C}_k&=&\frac{1}{3}[1-\Sigma_k\Sigma'_k+\frac{3}{2}k^2
\Sigma_k^{\prime 2}]\nonumber\\
\overline{F}_k&=&\frac{1}{6}[-4\Sigma_k^2+5k^2\Sigma_k\Sigma'_k+k^2]
\end{eqnarray}

$\tilde{\cal C}$ coefficients are defined as
\begin{eqnarray}
\tilde{\cal C}_2&=&2\int d\tilde{k}~
[\tilde{A}_k
(2X_k^3+2\frac{X_k^2}{\Lambda^2}+\frac{X_k}{\Lambda^4})+
\frac{k^2}{2}\Sigma_k^{\prime 2}(\frac{X_k}{\Lambda^2}+X_k^2)
]\nonumber\\
\tilde{\cal C}_3&=&\int d\tilde{k}~
[\tilde{B}_k(2X_k^3+2\frac{X_k^2}{\Lambda^2}+\frac{X_k}{\Lambda^4})
-(1-\frac{1}{2}k^2 \Sigma_k^{\prime 2})
(\frac{X_k}{\Lambda^2}+X_k^2)]\nonumber\\
\tilde{\cal C}_9&=&\int d\tilde{k}~
[\frac{1}{3}(k^2\Sigma'_k-\Sigma_k)\Sigma_k''X_k+
\tilde{D}_k(\frac{X_k^2}{\Lambda^2}+X_k^2)
+\tilde{E}_k(2X_k^3+2\frac{X_k^2}{\Lambda^2}+\frac{X_k}{\Lambda^4})
]\nonumber\\
\tilde{\cal C}_{10}&=&4i\int d\tilde{k}~
[\tilde{F}_k(2X_k^3+2\frac{X_k^2}{\Lambda^2}+\frac{X_k}{\Lambda^4})
+\frac{k^2}{2}\Sigma_k^{\prime 2}(\frac{X_k}{\Lambda^2}+X_k^2)]
\nonumber\\
\tilde{\cal C}_{11}&=&-2\int d\tilde{k}~k^2\Sigma_k'(\frac{X_k}{\Lambda^2}
+X_k^2)
\end{eqnarray}
with
\begin{eqnarray}
\tilde{A}_k&=&\frac{1}{3}(-2k^2\Sigma_k'+\Sigma_k)\Sigma_k(1
+2\Sigma_k\Sigma'_k)+\frac{1}{6}k^2(2\Sigma_k^2+k^2)
(\Sigma^{\prime 2}+\Sigma\Sigma'')-\frac{1}{3}\Sigma^2_k
+\frac{1}{6}k^2\Sigma_k\Sigma'_k
\nonumber\\
\tilde{B}_k&=&\frac{1}{3}\Sigma_k(-2k^2\Sigma'_k+\Sigma_k)(1+2\Sigma_k\Sigma'_k)+\frac{k^2}{18}(6\Sigma_k^2+k^2)(\Sigma^{\prime 2}+\Sigma_k\Sigma_k'')
\nonumber\\
&&-\frac{k^2}{6}(1+2\Sigma_k\Sigma'_k)-\frac{1}{3}(k^2+\Sigma_k^2
+k^2\Sigma_k\Sigma_k')]\\
\tilde{D}_k&=&-\frac{1}{2}k^2\Sigma^{\prime 2}+k^2(-\frac{1}{6}\Sigma_k\Sigma_k'+\frac{2}{9}k^2\Sigma_k'\Sigma_k'')(1+2\Sigma_k\Sigma_k')
+k^2\Sigma_k'(\frac{2}{9}k^2\Sigma_k'-\frac{1}{3}\Sigma_k)(\Sigma^{\prime 2}_k+
\Sigma_k\Sigma_k')]\nonumber\\
\tilde{E}_k&=&k^2\Sigma_k'(\frac{1}{6}\Sigma_k-\frac{1}{9}k^2\Sigma_k')(1+2
\Sigma_k\Sigma_k')^2\nonumber\\
\tilde{F}_k&=&\Sigma_k\Sigma_k'(\frac{1}{2}k^2-\frac{4}{3}k^2\Sigma_k\Sigma'_k+
\frac{2}{3}\Sigma^2_k)+k^2(\frac{1}{3}\Sigma_k^2+\frac{1}{9}k^2)
(\Sigma_k^{\prime 2}+\Sigma_k\Sigma_k'')-\frac{k^2}{3}(1+2\Sigma_k\Sigma'_k)+
\frac{1}{3}k^2]\nonumber
\end{eqnarray}

\end{document}